\documentclass[12pt,letterpaper]{article}
\usepackage{amsfonts,amssymb,amsmath,bbm,euscript,mathrsfs}
\usepackage[usenames]{color}
\usepackage{color,soul}
\usepackage[utf8]{inputenc}
\usepackage[T1]{fontenc}
\usepackage{textcomp}
\usepackage{graphicx}
\usepackage[tight,sf]{subfigure}
\usepackage{hyperref}

\newcommand{\sech}{\mathrm{sech}}

\newcommand{\sign}{\mathrm{sign} \,}

\title{\Huge Force dipoles and stable local defects on fluid vesicles}
\author{\Large
Jemal Guven${}^1$\footnote{\href{mailto:jemal@nucleares.unam.mx}{
jemal@nucleares.unam.mx}}
\, and Pablo
V\'azquez-Montejo${}^{1,2}$\footnote{\href{mailto:pvazquez@correo.cua.uam.mx}{
pvazquez@correo.cua.uam.mx}}
}
\date{}

\begin{document}

\maketitle

\begin{center}
{\it
$1$ Instituto de Ciencias Nucleares, Universidad Nacional Autónoma de México\\
 Apdo. Postal 70-543, 04510 M\'exico D.F., M\'exico\\
$2$ Departamento de Matemáticas Aplicadas y Sistemas, Universidad Autónoma
Metropolitana – Cuajimalpa,\\ C.P. 01120, México D.F., M\'exico}
\end{center}

\begin{abstract}
An exact description is provided of an almost spherical fluid vesicle with a fixed area and a fixed
enclosed volume  locally deformed by external normal forces bringing two nearby points on the
surface together symmetrically. The conformal invariance of the two-dimensional bending energy  is
used to identify the distribution of energy as well as the stress established in the vesicle.
While these states are local minima of the energy, this energy is degenerate;  there is a zero mode in the energy fluctuation spectrum, associated with area and
volume preserving conformal transformations, which breaks the symmetry between the two points.  The
volume constraint fixes the distance $S$, measured along the surface, between the two points; if it
is relaxed,  a second zero mode appears, reflecting the independence of the energy on $S$; in the
absence of this constraint a pathway opens for the membrane to slip out of the
defect. Logarithmic curvature singularities in the surface geometry at the points of contact signal
the presence of external forces.  The magnitude of these forces varies inversely with $S$ and so
diverges as the points merge; the corresponding  torques vanish in these defects.  The geometry
behaves near each of the singularities as a biharmonic monopole, in the region between them as a
surface of constant mean curvature, and in distant regions as a biharmonic quadrupole. Comparison of the distribution of stress with the quadratic approximation in the height functions points to shortcomings of the latter representation. Radial  tension is accompanied by lateral compression, both near the singularities and far away, with a crossover from tension to compression occurring in the region between them.
\end{abstract}

PACS: 87.10.Pq, 87.16.D-, 02.40.Hw

\section{Introduction}

\begin{figure}[htb]
\begin{center}
\includegraphics[scale=0.5]{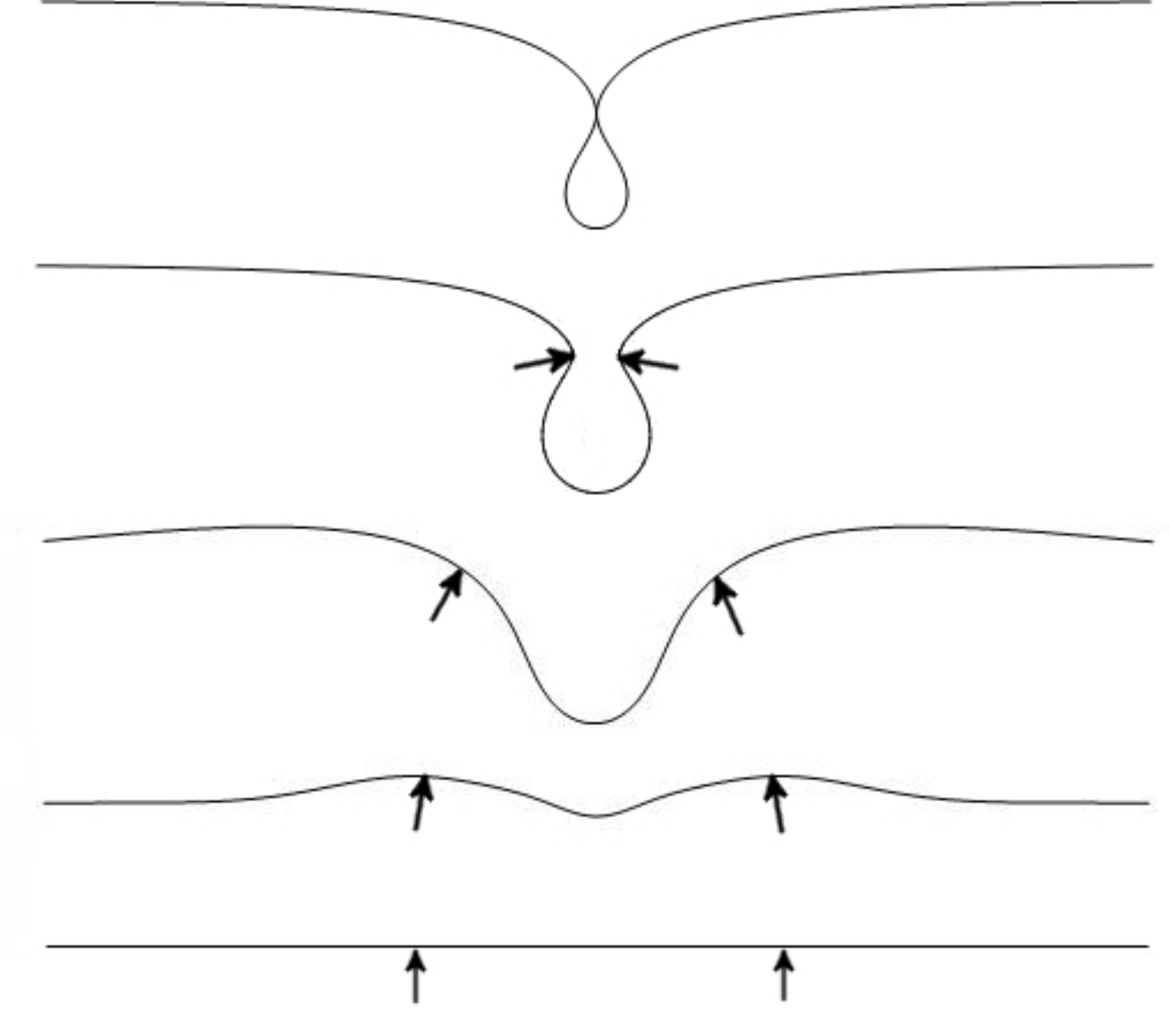}
\end{center}
\caption{\small{Sequence of normal deformations bringing two points on an almost spherical membrane
together. }}
\label{fig1}
\end{figure}

\begin{figure}[htb]
\begin{center}
 \subfigure[]{\includegraphics[clip, trim= 0.5cm 7cm 0.5cm 0.5cm, scale=0.21]{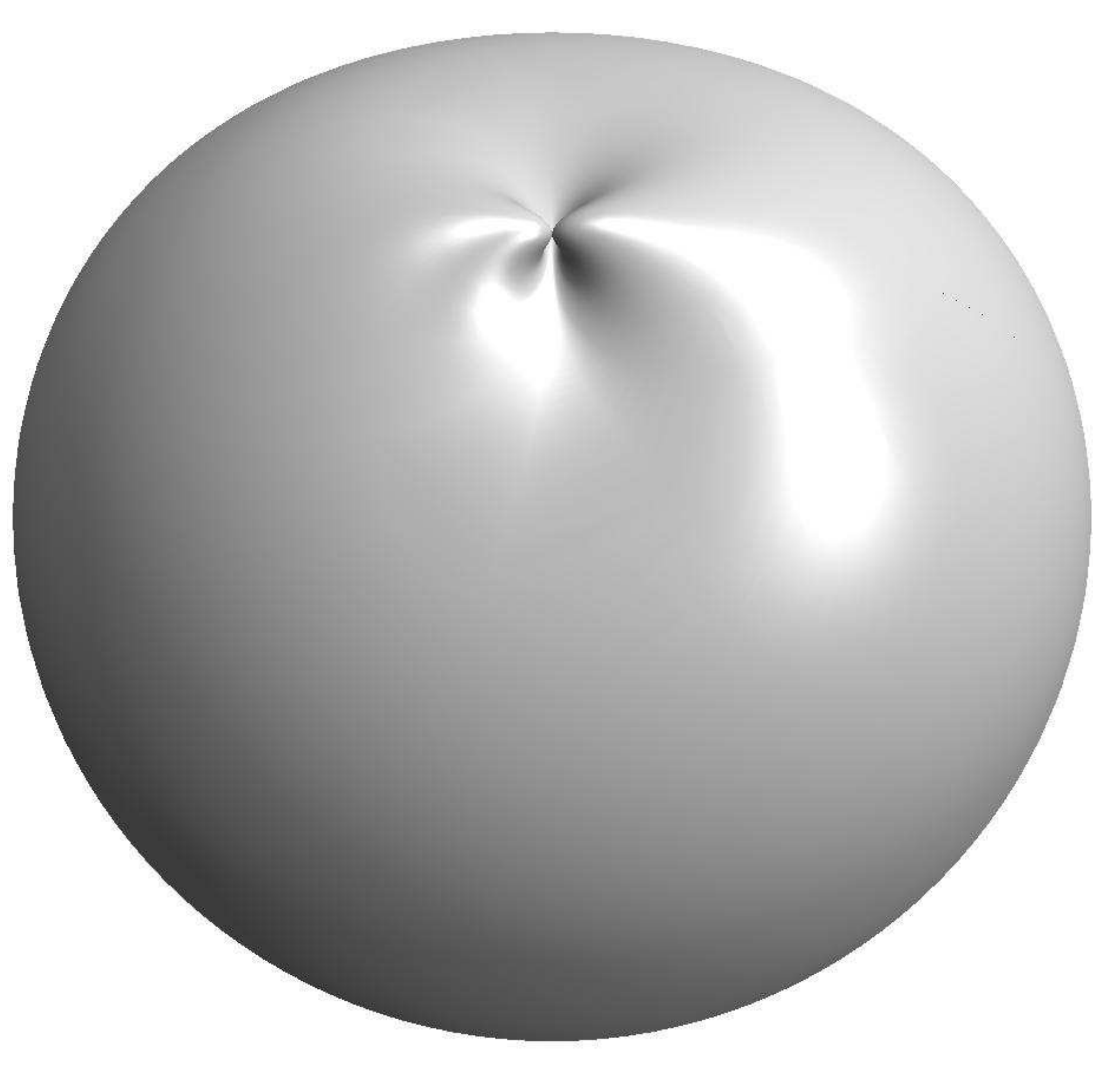}}
 \hfil  \subfigure[]{\includegraphics[scale=0.11]{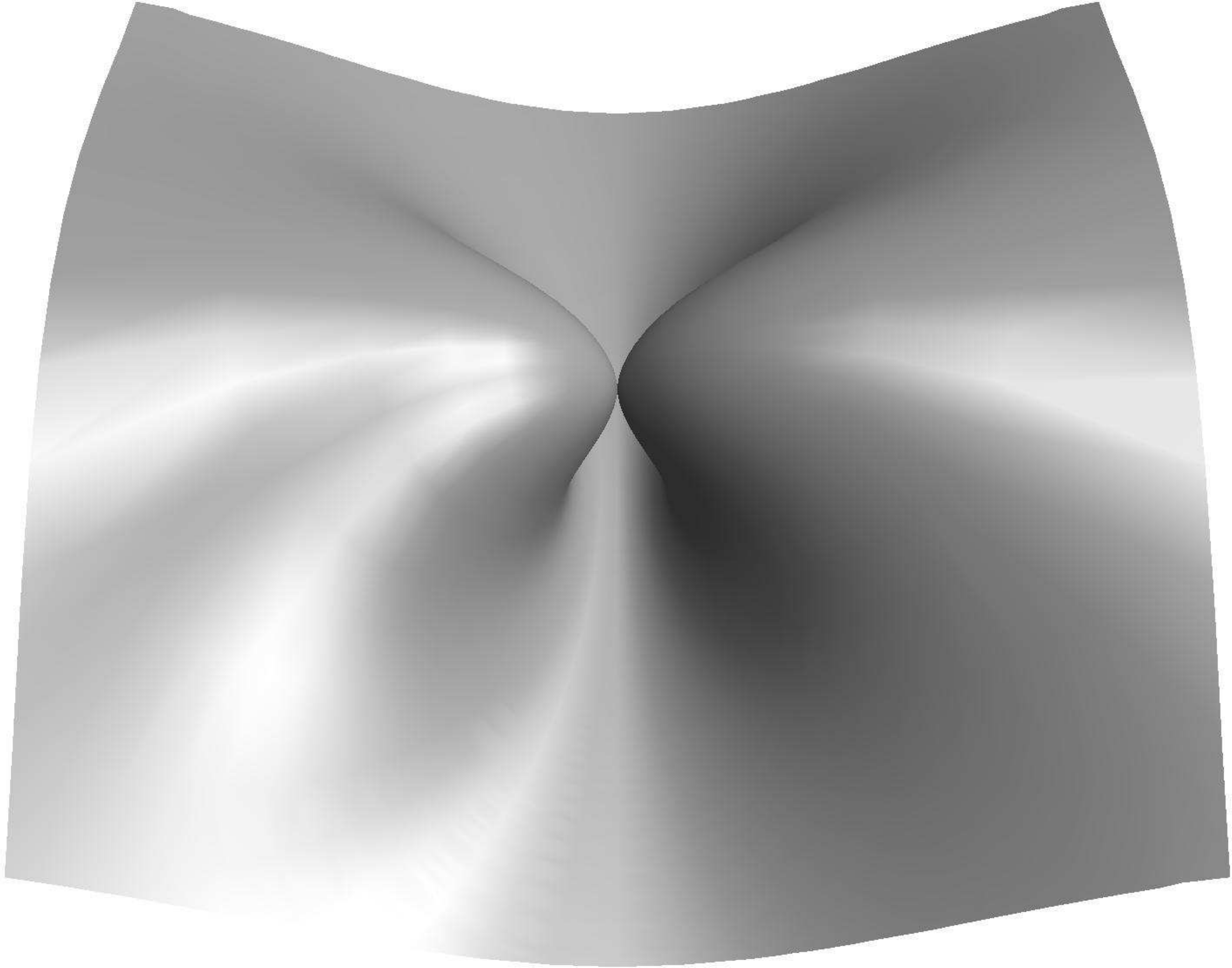}}\\
 \subfigure[]{\includegraphics[clip, trim= 0.5cm 5cm 0.5cm 0.5cm, scale=0.21]{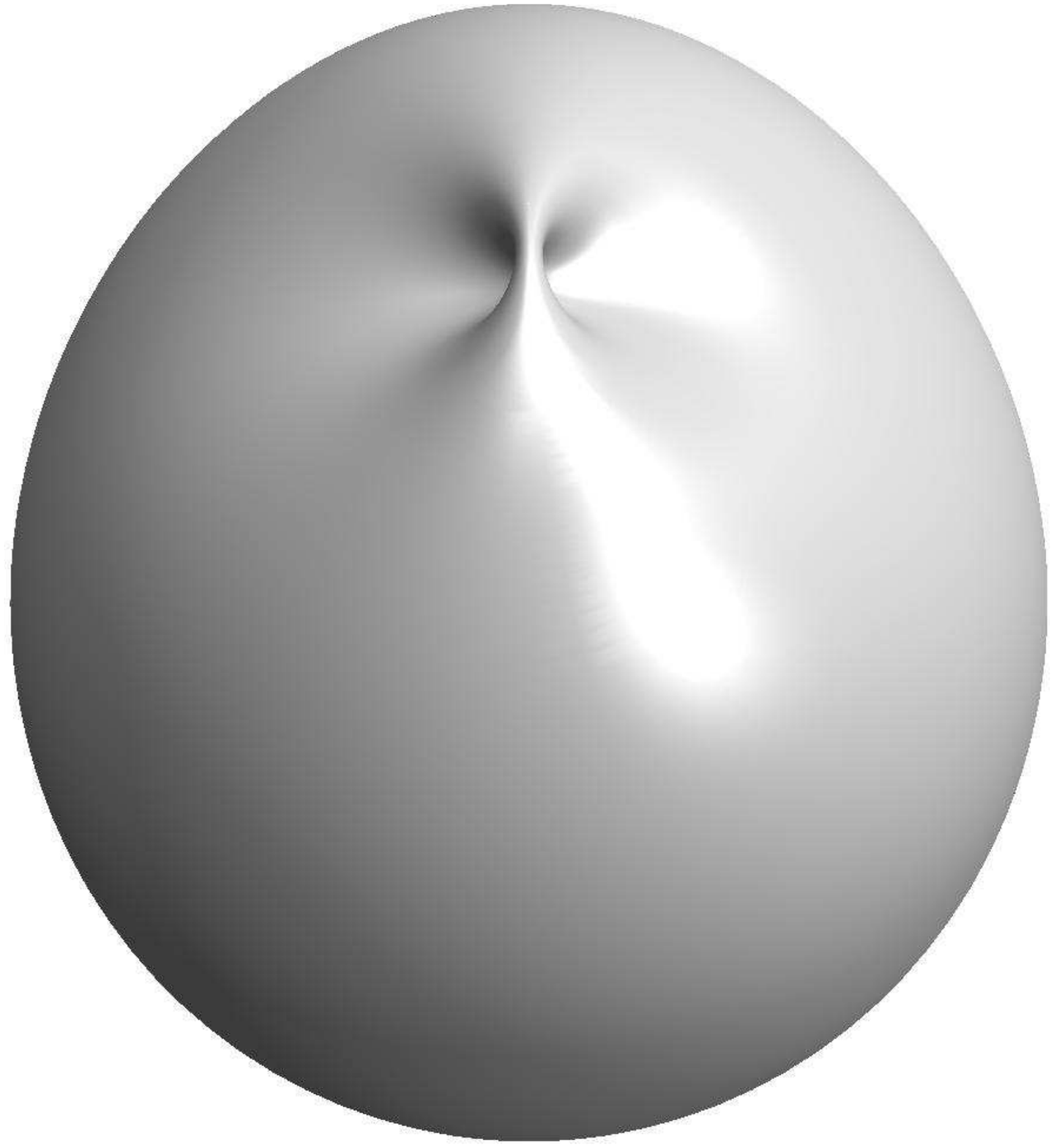}}
\hfil \subfigure[]{\includegraphics[scale=0.11]{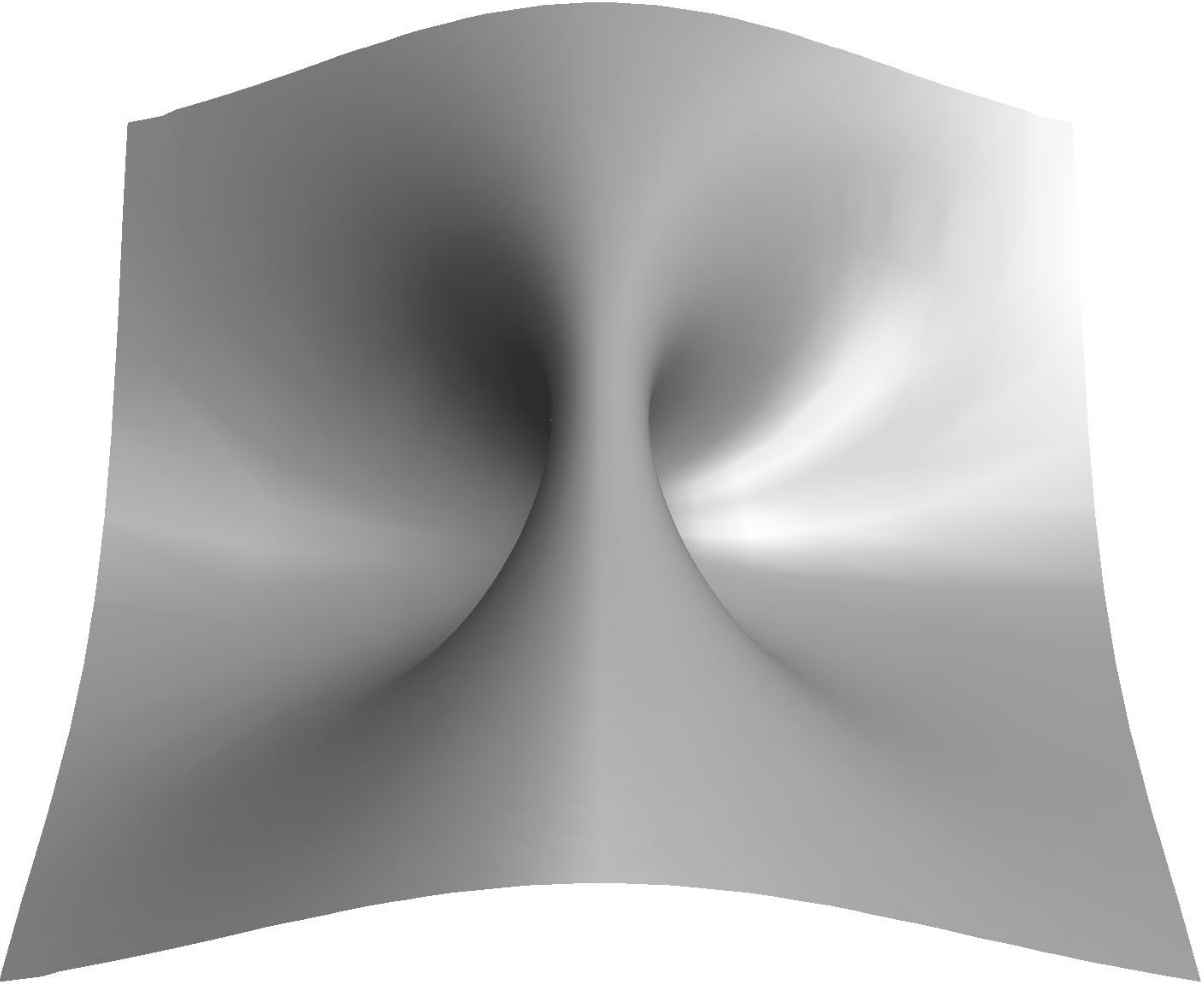}}
\end{center}
\caption{\small Geometry of localized defects on an almost spherical vesicle. (a) Membrane with
two touching fingers pushed out. (b) Close-up of the fingers. (c) Membrane pinched  into self-contact. (d) Close-up of the pinch. These two surfaces are obtained with the center of inversion placed at ${\bf x}_0= 1.07 \,{\bf i}$ and ${\bf x}_0 =0.93\, {\bf i}$, respectively, as explained in Sec. \ref{sectinvknoids}.} \label{fig2}
\end{figure}

A striking feature of cellular membranes in biology is that they behave as fluid membranes on
mesoscopic length scales.  Their mechanical properties are captured by the geometrical degrees of
freedom of the membrane surface, which is described with extraordinary accuracy by the
Canham-Helfrich energy, quadratic in the curvature \cite{canhamHelfrich, Willmore},
\begin{equation} \label{CHenergy}
 H_{CH} = \kappa H_1+ {\bar \kappa} H_{GB} + \sigma (A-A_0) +
P (V-V_0) \,,
\end{equation}
where
\begin{equation}
H_1[{\bf X}] = \frac{1}{2}\, \int dA\, K^2 \,,\quad H_{GB} [{\bf X}] =  \int dA \, K_G \,;
\end{equation}
where $K$ is twice the mean curvature, $K_G$ the Gaussian curvature of the surface, given by the
trace and determinant of the curvature tensor, $K=C_1+ C_2$, and $K_G= C_1 C_2$ with $C_1$ and
$C_2$ the principal curvatures. $H_{GB}$ is the Gauss-Bonnet topological invariant. The physical
parameters $\kappa$ and $\bar{\kappa}$ are the bending and saddle splay modulus, with units
of energy. The remaining two parameters $\sigma$ and $P$ are  Lagrange multipliers fixing the
membrane area $A$ at some value $A_0$ and its enclosed volume $V$ at a value $V_0$.
\vskip1pc \noindent
While it is simple to write down the energy, the nonlinear field theory associated with it is not
simple. Thus, unless the membrane lends itself to a description in terms of a height function which
varies slowly above some simple geometry, it will tend to be necessary to examine its behavior
numerically.
\vskip1pc \noindent
On occasion, the symmetries of the theory allow one to circumvent this necessity. In this paper, a
stratagem exploiting the conformal symmetry of the bending energy will be used to provide an exact
analytical description of a local deformation that is established in an almost spherical vesicle when
two neighboring points are brought into contact.
\vskip1pc \noindent
Consider two closely separated points on an almost spherical vesicle. The obvious way
to bring the two together is to apply tangential forces. In the geometric approximation there is
no penalty associated with surface shear on a fluid membrane, and, thus the two points can be brought
together this way without deforming the membrane.  It is also possible, however, to apply  forces
normal to the surface,  with appropriate torques, in such a way that the points remain
separated on the surface but are brought together in space. An illustration of this process is
shown in Fig. \ref{fig1}. These forces can be directed in or out of the vesicle.
\vskip1pc \noindent
Views of the corresponding deformed vesicles at the end point of this process are shown in
Figs. \ref{fig2}(a) and \ref{fig2}(c). We show a close-up of the local behavior of the vesicle in Figs.
\ref{fig2}(b) and \ref{fig2}(d).
\vskip1pc \noindent
The surprise is that, even though the membrane undergoes large deformations in a
neighborhood of the two points under the influence of the force, an exact description of this final
state exists. In this state, the external forces are horizontal, and the torques about the point of
contact vanish. This description is a happy accident of the conformal invariance of the two-dimensional bending energy  \cite{Willmore,Polyakov,Seifert}.  In contrast,  neither the initial state, nor any of the intermediate states in the
sequence before contact is made, involving nontrivial external forces and torques, lend themselves
to an analogous treatment. While this state may represent an idealization from a physical point of
view, and developing an experimental protocol to construct it will provide its own
challenges, it is nonetheless a genuine nonperturbative feature of the mesoscopic description of
fluid membranes, and it does provide unexpected access to the nonlinear response of fluid membranes
to external forces.
\vskip1pc\noindent
Constraints would typically be expected to break the full conformal invariance of the bending energy.
However,  the contact constraint associated with the end state is invariant under conformal transformations (in particular, it does not introduce an additional length scale).  Fixing the area will restrict  the invariance group to the subgroup of area conformal transformations. As we will see,  conformal transformations preserving the area of the equilibrium surface are described by two parameters.\footnote{In general, a three parameter family would be expected; this is reduced by the uniqueness, modulo scale, of the catenoid as an axially symmetric minimal surface.}
Likewise, their counterparts preserving volume will be characterized by two-parameters.
Simultaneously constraining the area and the volume restricts the invariance group to a smaller subgroup: characterized by a single parameter. A variational argument showing that the multipliers vanish under these circumstances is provided in \ref{appCIC}.
\vskip1pc \noindent
Axially symmetric membrane states with two poles brought together were constructed a few years ago
\cite{guvencastro, invcat}. This was done by exploiting the fact that any state related by inversion
in a sphere to any equilibrium state will also be an equilibrium state. In particular,  states
described by inversion of a catenoid--an axially symmetric infinite cylindrical surface growing
exponentially at its two ends--are also equilibria. Under inversion, the topology of the catenoid
will be changed as its two ends get compactified to a single point. The compact axially symmetric
vesicle states described in Refs. \cite{guvencastro, invcat} result when the center of inversion lies
on the axis of symmetry. By positioning this point instead close to the surface of the
catenoid in the neighborhood of its neck,  nearby points on the surface get inflated into a large
spherical region. A very different closed vesicle is obtained: spherical almost everywhere, with a localized
defect--held together by a force dipole--illustrated in Fig. \ref{fig2}, sitting at one of its poles.
\vskip1pc \noindent
Any one of these equilibrium configurations will have a particular
area, $A_0$, and a particular volume, $V_0$.  Since it is a stationary state of
the unconstrained energy, it will also be a stationary state of the
constrained energy in which  the area and volume are constrained to be $A_0$
and $V_0$.  By construction, it will satisfy the new constraints.
\vskip1pc \noindent
If the center of inversion lies just outside the catenoid, the local geometry of the  defect
resembles a flat sheet with two fingers of membrane--touching at a point-- pushed up from below, a
valley forming beneath them, as illustrated in Figs. \ref{fig2}(b).  If, on the other hand, the
point of inversion lies just inside,  the exterior resembles a flat sheet that has been pinched into
self-contact; a ridge forms above the two points that have been brought into contact,  illustrated
in Fig. \ref{fig2}(d), not unlike the wrinkle that is formed in skin that has been pinched together
between two fingers. Locally, these two geometries are identical; one is the
other viewed from below: The interior of the defect represented in Fig. \ref{fig2}(a) is
indistinguishable from that represented in Fig. \ref{fig2}(c).
\vskip1pc \noindent
A feature of this geometry is that a well defined tangent plane exists at the two points of
contact.  This is consistent with a logarithmic singularity in the curvature at these two points. In
this construction, these singularities  originate in the compactification of the exponential ends
\cite{guvencastro, invcat}. The asymptotic behavior of the minimal surface thus completely determines the
geometry at the center of the defect.  The conformal symmetry of the bending energy provides
a connection between the weak-field harmonic behavior of the catenoid at infinity, described
accurately by perturbation theory, and the nonlinear behavior of the defect at its center. This
correspondence is a general feature of field theories exhibiting an underlying conformal symmetry.
\vskip1pc \noindent
The focus here, for simplicity,  will be on mirror symmetric states, generated by placing the point of inversion on the mirror plane containing the waist of the catenoid. The scale is set by fixing the area; finger-like and pinch-like almost-spherical symmetric states are characterized uniquely by the geodesic distance $S$
separating the two points along the surface which, in turn we will see, is determined by the fixed volume. Thus, if both area and volume are fixed, there is a unique symmetric state of each kind.\footnote {The two points  will  migrate along the surface in the process of bringing them together. Once equilibrium is established, however, the distance between them will  be set by the constraints, independent of their initial  positions on the surface.}
In a future publication, the breaking of the mirror symmetry will be explored \cite{Trinoid}.
\vskip1pc\noindent
Even though the total energy will be independent of the distance separating them along the surface,
the  degree of localization of the energy density will depend on this distance. What is more, we will show that if the forces and torques are fixed, this state will be energetically stable, whether the perturbation breaks the symmetry or not. This means that there are no negative eigenvalues in the fluctuation spectrum. Unless the configuration is a sphere, it will have a different ratio of $V_0^2/A_0^3$ than a sphere does, so it cannot be transformed into a sphere by energy, volume, and area preserving transformations. There will, however, be a zero mode of the energy associated with conformal transformations that preserve both the vesicle area and its volume. This zero mode
breaks the mirror symmetry between the two poles. If the constraint on the volume is relaxed,  a second zero mode appears. For symmetric states, this corresponds to changing the geodesic distance between the poles.
Under these circumstances, there is  a path connecting the defect state to a sphere with $S = 0$,
the same energy and the same area; because the defect state has the same topology as a sphere, it
is  not protected  topologically. Such a mode would allow the membrane to slip out of the defect.
\vskip1pc \noindent
The normal force tying the points together provides a source of stress in the membrane.\footnote
{The catenoid generating this geometry is itself a stress-free equilibrium state of a fluid
membrane, albeit one that is never realized physically. This is not a problem here for its only role
will be to generate the state of physical interest.}  If these forces were removed, the membrane sheet would
collapse into one of the equilibria of a fluid membrane described, for axially symmetric vesicles, in Refs. \cite{SvetinaZeks} and \cite{SeifertBerndlLipowsky}. In the absence of the volume constraint this would be
a featureless spherical state, the only unconstrained  equilibrium geometry with this topology. It is these forces that establish the defect geometry and, thus, the tangent planes along which the molecules forming the fluid membrane may flow. In the absence of an appropriate zero mode, the vesicle cannot slip out of the defect.\footnote {Other symmetric states with two points held together  may exist. They will, however, involve more complicated applied forces and non-vanishing torques.}
\vskip1pc \noindent
The stress finds itself concentrated in the neighborhood of the two points; it is also
completely captured by the defect geometry. The curvature singularities in this geometry are
associated with these localized external forces and manifest themselves as divergences in the
stress.  The strength of these forces is registered by an appropriate integral of the stress along a
contour encircling one of the points. As we will demonstrate, it is also possible to evaluate them
exactly by deforming the contour to exploit the symmetry of the problem. Unlike the energy, they vary
inversely with the distance $S$ between the two points measured along the surface.
The same argument can be tweaked to show that the external torques about these points vanish.
\vskip1pc \noindent
Conformal invariance also permits an exact analytic description of the inhomogeneous and anisotropic
distribution of stress established in the membrane in the fully non-linear theory. This is the first
instance--to our knowledge--of such a description  in any non-trivial fluid membrane geometry.
Three qualitatively different regimes are identified: the neighborhood of the singularities within the
fingers, the valley that extends under them, and the far region. While a global description
of the surface in terms of a height function does not exist, in each of these three regions it is
possible to describe the surface in terms of its height above an appropriate plane, and to compare
the exact stress tensor with its counterpart in the  small gradient biharmonic approximation. In
particular, it will be shown that, whereas  near the singularities the geometry behaves as a
biharmonic monopole, in the valley it behaves as a surface of constant mean curvature, and
far away it behaves as a biharmonic quadrupole.  Characteristic distributions of stress are
found to be associated with each region; both near the poles as well as asymptotically,  radial
tension is accompanied by an equal lateral compression. A crossover from tension to compression is
encountered in the region between the poles.

\section{Geometry of the Defect} \label{sectinvknoids}

Consider a catenoid, with unit neck radius, aligned along the $Z$ axis.\footnote{Three-dimensional
Euclidean space will be described by Cartesian coordinates  $X,Y,Z$, with basis vectors ${\bf
i}=(1,0,0)$, ${\bf j}=(0,1,0)$ and ${\bf k}=(0,0,1)$. Polar coordinates $\rho$ and  $\varphi$  are
adapted to the $X-Y$ plane.} Its polar radius $\rho$ at a given value of $Z$  is given by
\cite{Gray, Fomenko} $\rho(Z) = \cosh{Z}$, so the surface is described parametrically  in the
form
\begin{equation} \label{axysymcatenoid}
{\bf X}_0 (Z,\varphi) = \left(\rho (Z) \cos \varphi,\rho (Z) \sin \varphi,Z \right)\,.
\end{equation}
\begin{figure} [htb]
\begin{center}
\subfigure[Catenoid]{\includegraphics[scale=0.15]{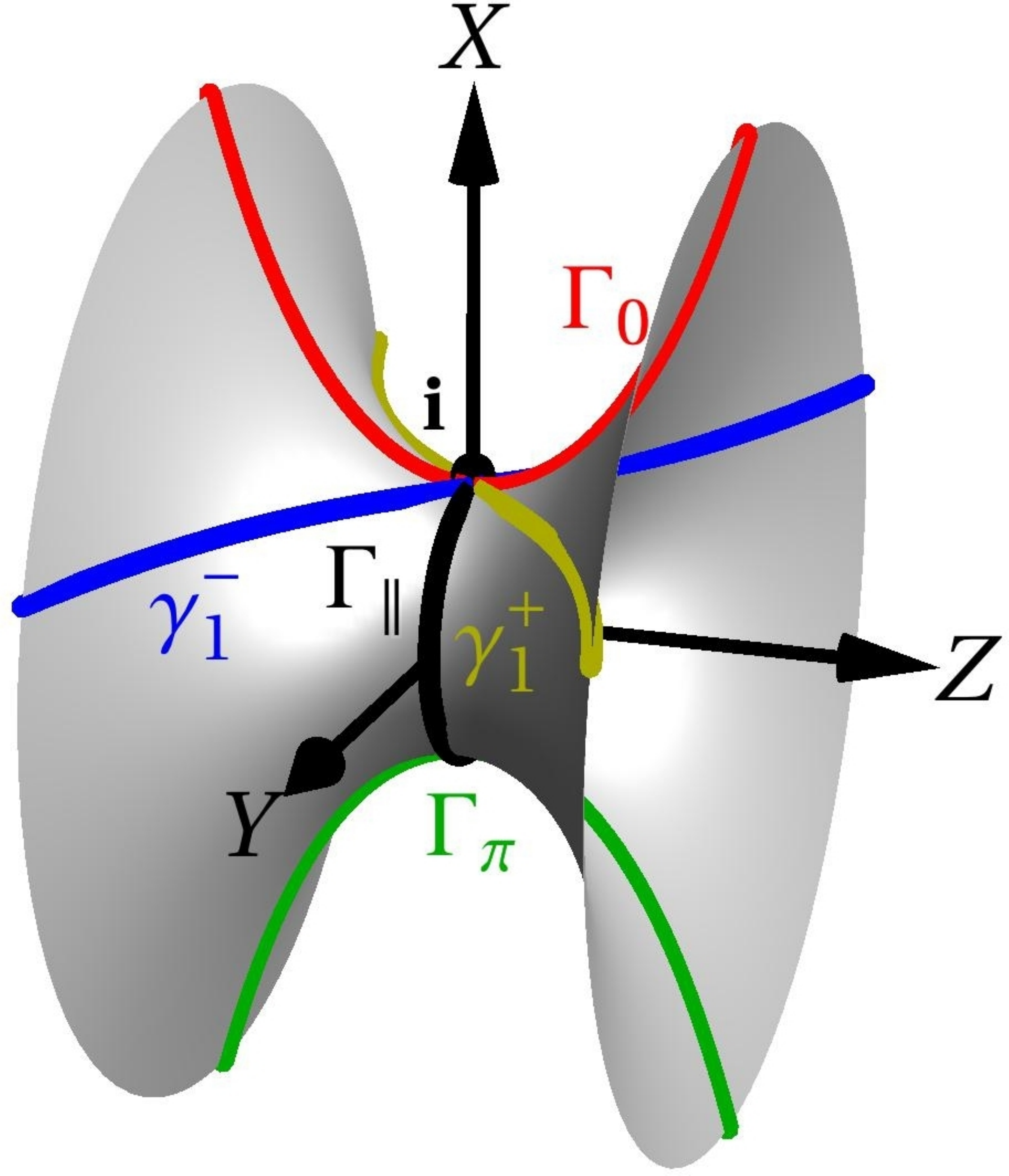}} \qquad
\subfigure[Defect]{\includegraphics[scale=0.14]{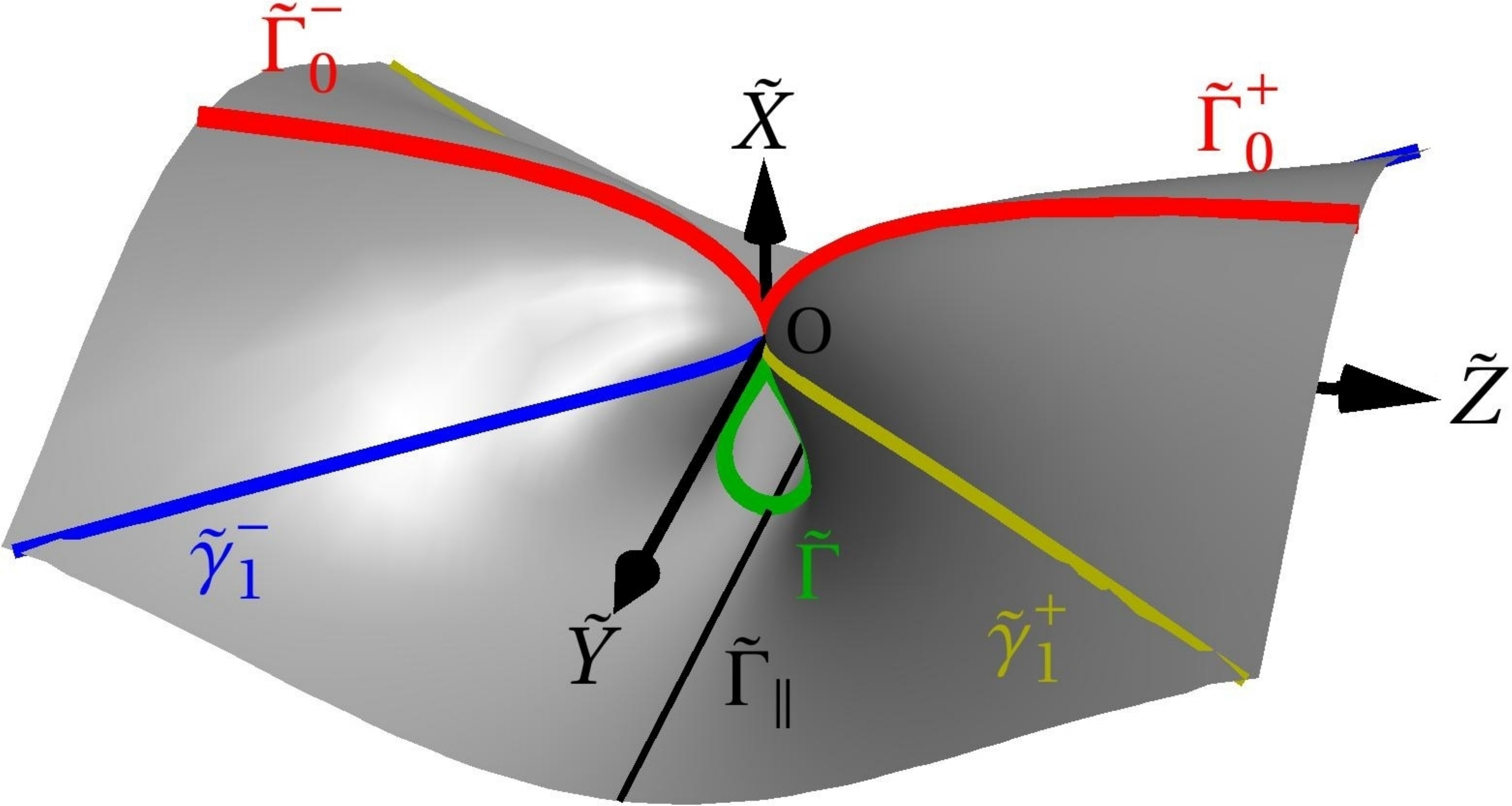}}
\caption{\small (Color online) (a) Surface curves on the catenoid obtained from its
intersection with relevant planes: catenaries on the plane $Y=0$
with $\varphi=0$ ($\Gamma_0$); $\pi$ ($\Gamma_\pi$) are represented
by red (upper) and green (lower), curves respectively, the circle
$\Gamma_\|$ on the plane $Z=0$ with a black line; and curves on the plane $X=1$ ($\gamma_1^\pm$)
with blue (front left) and yellow (front right) lines. The center of inversion is
indicated by the
point ${\bf i}$. (b) The image under inversion in a unit sphere
centered at ${\bf i}$. } \label{fig3}
\end{center}
\end{figure}
The surface obtained by inversion of this catenoid in a sphere of radius $R$, centered at the point
${\bf i}$ on its neck, will be described by the new position vector $\widetilde{\bf X} = R^2 {\bf
X}/ |{\bf X}|^2$, where ${\bf X} = {\bf X}_0-{\bf i}$, \cite{Gray}, given explicitly
by\footnote{This surface is also parametrized by $Z$ and $\varphi$.
${\bf X}$ is dimensionless in our construction. This implies that $R^2$ must have dimensions of
length in order for the physical geometry described by $\widetilde{\bf X}$ to possess dimensions of length.}
\begin{equation}
\label{eq:inv} \widetilde{\bf X} = \frac{R^2}{Z^2 + P^2(\rho,\varphi)} \, \left(\rho \cos \varphi-1,\rho \sin
\varphi,Z \right)\,,  \qquad P^2(\rho, \varphi) = \rho^2 - 2 \rho \cos \varphi +1\,.
\end{equation}
The two planar ends of the catenoid, represented by the neighborhood of the points at infinity
$Z\rightarrow \pm \infty$ get compactified to a single point, the origin of the Euclidean plane,
$O$.  Because the center of inversion lies on the
catenoid, the neighborhood of this point is mapped to infinity.  Inversion generates another infinite surface.
This is the defect geometry illustrated in \ref{fig3}(b), a labeled reproduction of Fig.
\ref{fig2} (b) (or viewed below as Fig. \ref{fig2}(d)). This does not, of course, represent a defect
on a vesicle of fixed finite surface area.  But it is simple to tweak the construction to do so: by
moving the center of inversion off the catenoid,  its neighborhood on the catenoid
get compactified into a sphere.  If this point is located close to the neck, the vesicle geometry
will approximate  an almost round sphere  with the defect located at its north pole.
Geometries corresponding to a point of inversion lying both outside as well as inside the catenoid are
illustrated in Figs. \ref{fig2}(a) and \ref{fig2}(c). Notice  in particular that the distance between the
two points, as measured along the surface, is small compared to the radius of the sphere. As  we
will quantify below, the local defect geometry in an almost-spherical vesicle is accurately
described by a center of inversion lying on the catenoid itself.
\vskip1pc \noindent
To describe the defect geometry in greater detail, first note that its geometry possesses $\tilde Y=0$ and $\tilde Z=0$ as mirror planes: the former is what remains of the axial symmetry and the latter is the unbroken mirror symmetry of the catenoid. These two symmetries are captured respectively by $\varphi\leftrightarrow -\varphi$ and $Z\leftrightarrow -Z$. One can now trace the image under inversion of various surface curves on these two mirror planes:
\begin{enumerate}
\item Two catenaries describe the intersection of
the catenoid with the plane $Y=0$ (given by $\varphi =0 , \pi$).
\vskip1pc \noindent
The  catenary  with $\varphi =0$, $\Gamma_0$, passing through the center of
inversion ${\bf i}$, represented by the red (upper) curve in Fig. \ref{fig3}(a), is
mapped to two semi-infinite curves $\widetilde \Gamma_0^-$ and
$\widetilde \Gamma_0^+$ (with $Z<0$ and $Z>0$ respectively)
\begin{equation}
 \widetilde \Gamma^{\pm}_0 = \frac{R^2}{Z^2+\left(\rho-1\right)^2}
\left(\rho-1,0,\pm |Z|\right)\,,
\end{equation}
both of which terminate at $O$ and represented in red
in Fig.\ref{fig3}(b).
They both increase to an asymptotic value $\tilde X = R^2/2$.
\vskip1pc \noindent
The second catenary with $\varphi=\pi$, $\Gamma_\pi$, maps to the finite curve
$\widetilde \Gamma$ whose two ends also terminate at $O$. They are represented
by the green (lower) curves in Figs. \ref{fig3}(a) and \ref{fig3}(b) respectively. The point
$(-1,0,0)$ on the
catenary maps to the point $(-1/2,0,0)$ on this curve. Together $\widetilde
\Gamma_\perp = \widetilde \Gamma_0^{-} \cup \widetilde \Gamma \cup  \widetilde
\Gamma_0^{+}$ form a single curve with a continuous tangent vector,
making contact with itself at $O$.
\item The circular waist of the catenoid $\Gamma_\parallel$ resulting from
its intersection with the plane $Z=0$ (illustrated with a black
curve in Fig.\ref{fig3}(a)) passes through the center of inversion,  and thus
under inversion opens into the straight line $\tilde{\Gamma}_\parallel$
(black line in Fig.\ref{fig3}(b)) parametrized by
\begin{equation}
\tilde{\Gamma}_\parallel = \frac{R^2}{2} \left(-1,\cot (\varphi/2),0\right)\,,
\end{equation}
which intersects $\widetilde \Gamma$ orthogonally.
\item The curves $\gamma^{\pm}$ resulting from the intersection of the catenoid
with the plane $X=1$, given by $\tan \varphi = \pm \sinh Z$, get mapped to
the curves
\begin{equation}
\tilde{\gamma}^{\pm}_1 = \frac{R^2}{Z^2+\sinh^2{Z}} (0,\pm\sinh{Z},Z)\,,
\end{equation}
 which are the intersection of the defect surface with the plane
$\tilde{X}=0$. These two curves are represented by yellow (front right) and blue
(front left) lines in Figs. \ref{fig3}(a) and \ref{fig3}(b).
This plane represents the midplane of the asymptotic geometry, described in detail in
Sec. \ref{asymptgeom}.
\end{enumerate}
\vskip1pc\noindent
The curves $\widetilde \Gamma_\perp$ and $\widetilde \Gamma_\|$ are geodesics on the surface.

\subsection{Characterizing symmetric defects on  almost spherical vesicles}\label{character}

\begin{figure} [htb]
\begin{center}
\subfigure[]{\includegraphics[scale=0.66]{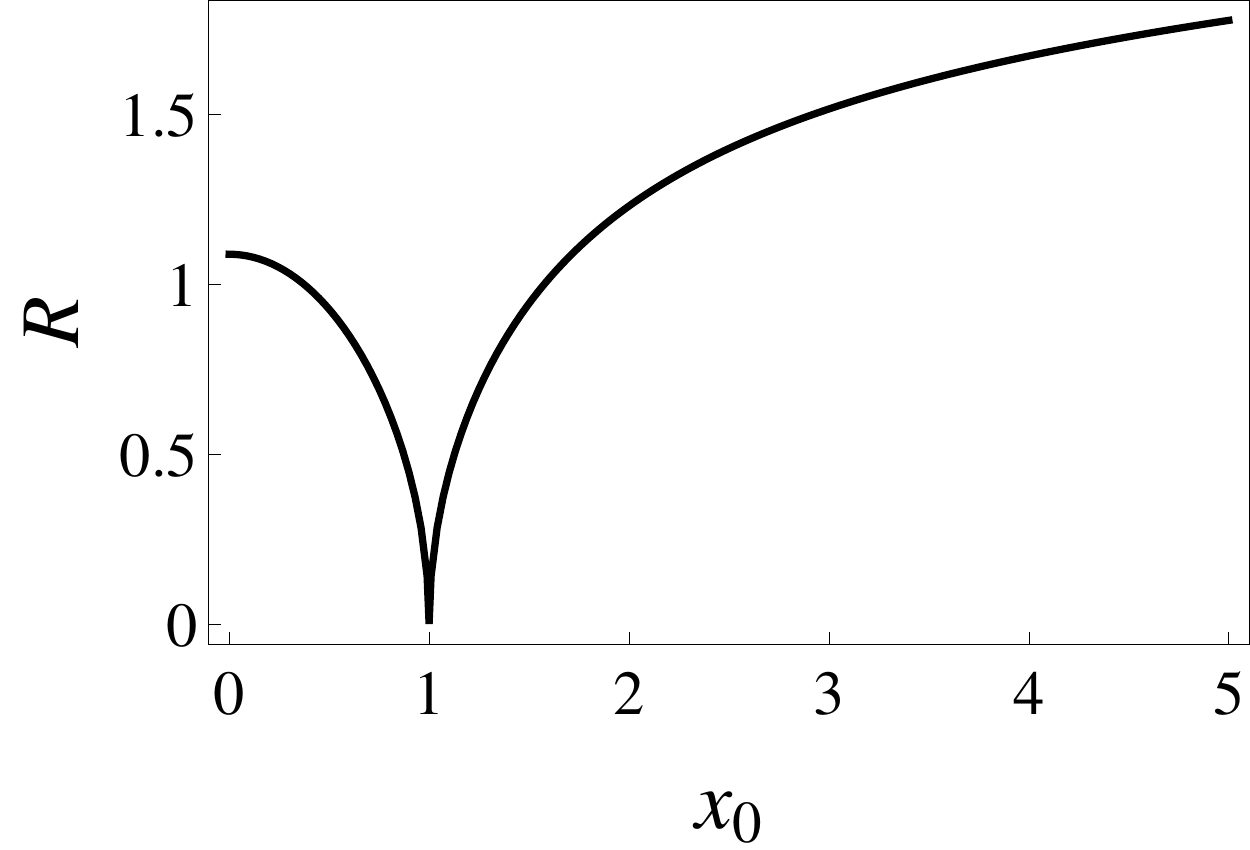}} \qquad
\subfigure[]{\includegraphics[scale=0.66]{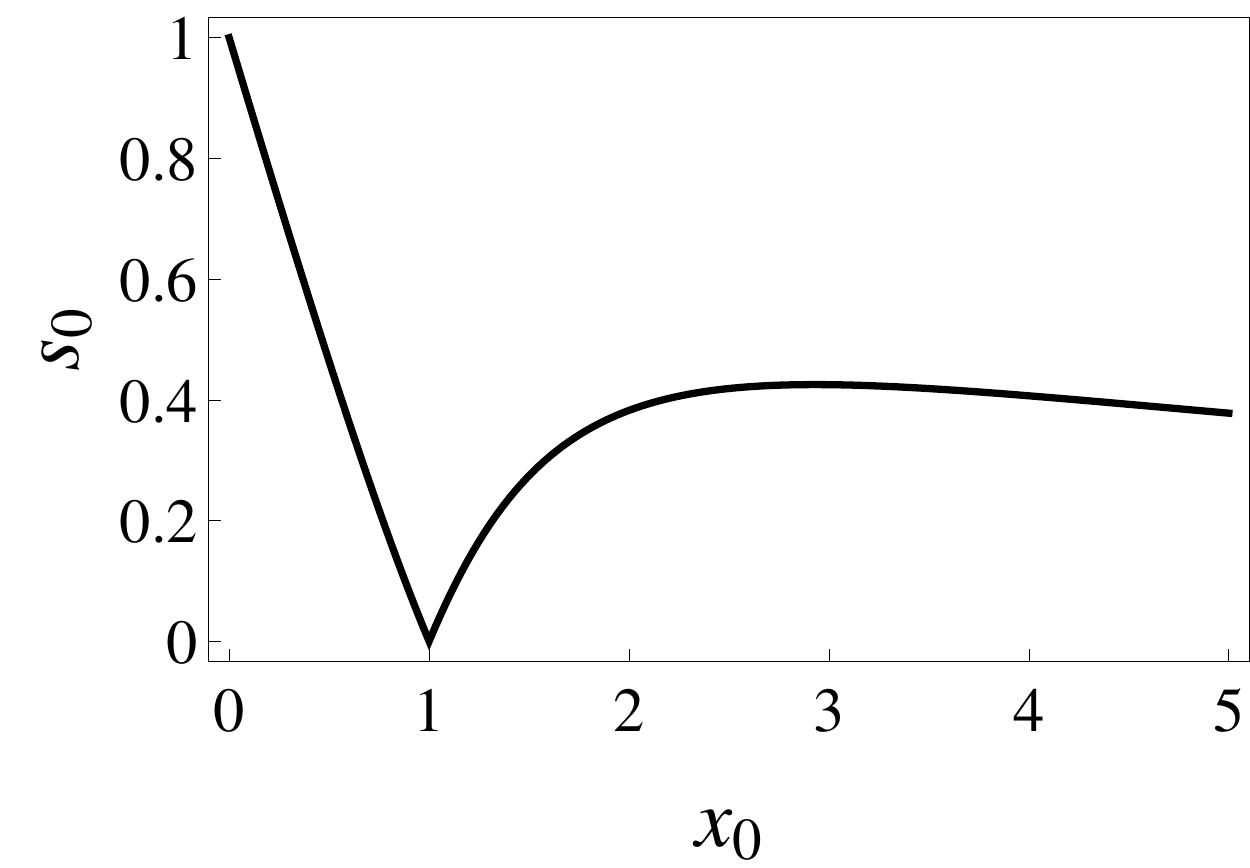}} \\
\subfigure[]{\includegraphics[scale=0.66]{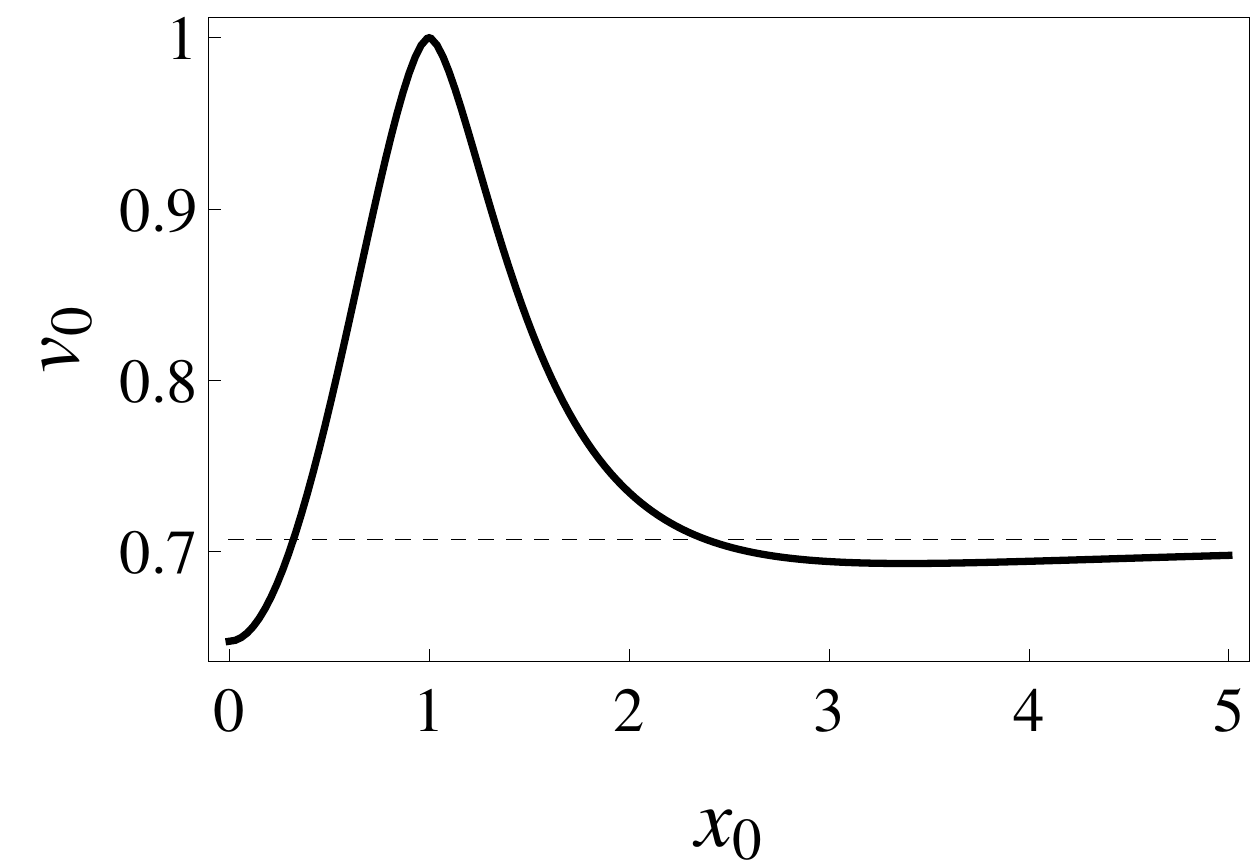}} \qquad
\subfigure[]{\includegraphics[scale=0.66]{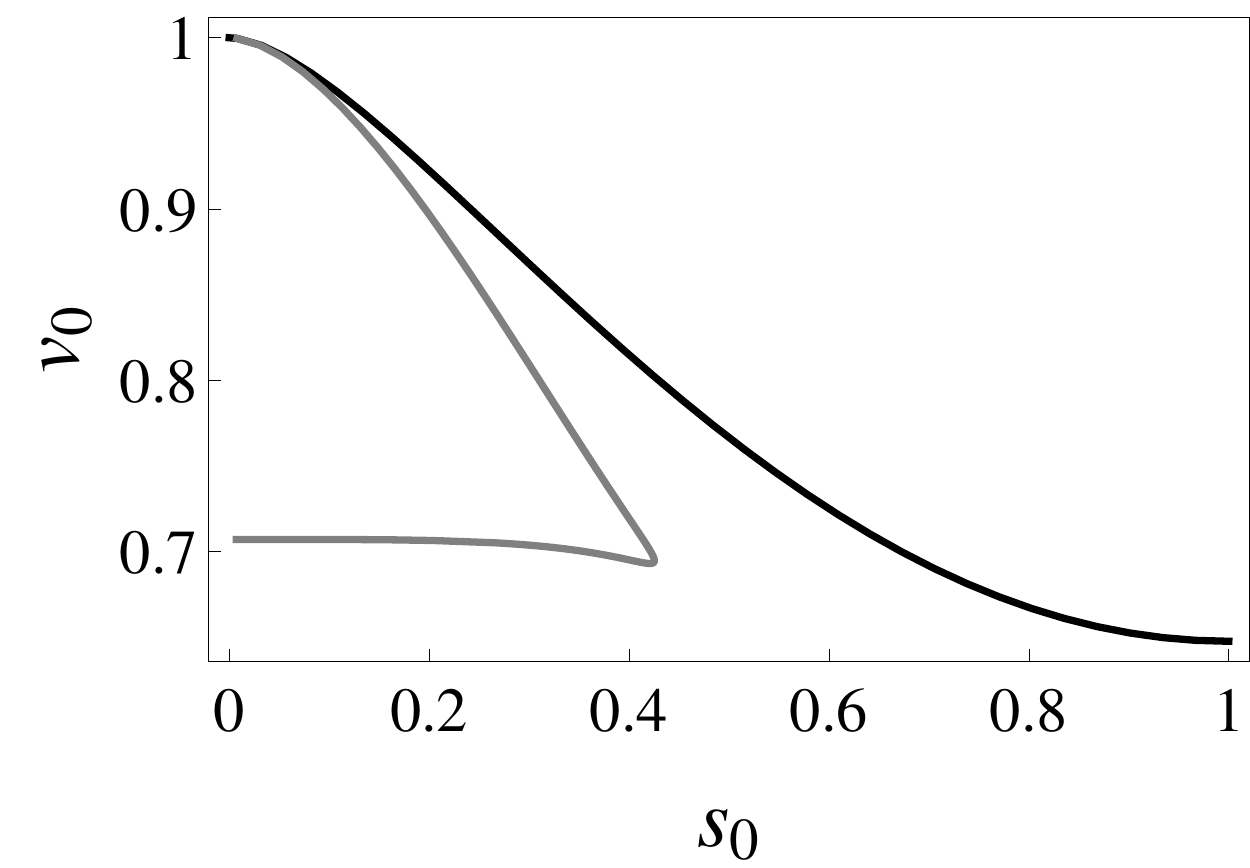}}
\caption{\small (a) Radius of inversion $R$ vs. center of inversion $x_0$; (b) normalized distance
between the defects $s_0=S/S(0)$ vs. $x_0$ ($S(0)=3.0669$ is the distance
between poles in symmetric discocyte); (c) Normalized enclosed volume $v_0=V_0/V_S$ vs. $x_0$
($V_S=4\pi/3$ is the volume of the unit sphere). $v_0$ tends asymptotically to the value
$1/\sqrt{2}$ as $x_0 \rightarrow \infty$;
(d) $v_0$ vs. $s_0$. The black (gray) line represents surfaces with $x_0 \leq 1$ ($x_0 > 1$).}
\label{fig4}
\end{center}
\end{figure}

\begin{figure} [htb]
\begin{center}
\subfigure[$x_0=0.5$]{\includegraphics[scale=0.1]{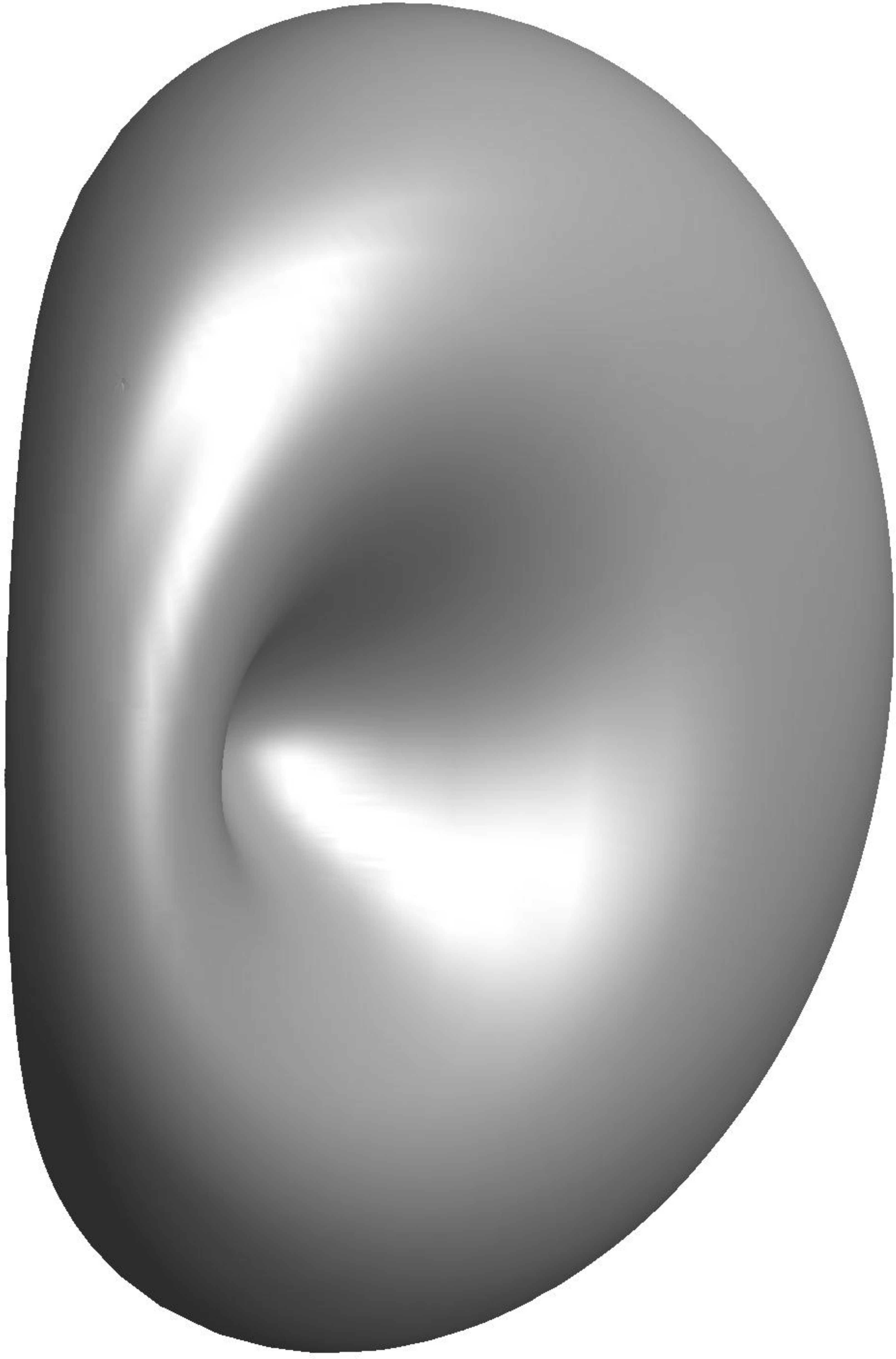}} \qquad
\subfigure[$x_0=2$]{\includegraphics[scale=0.14]{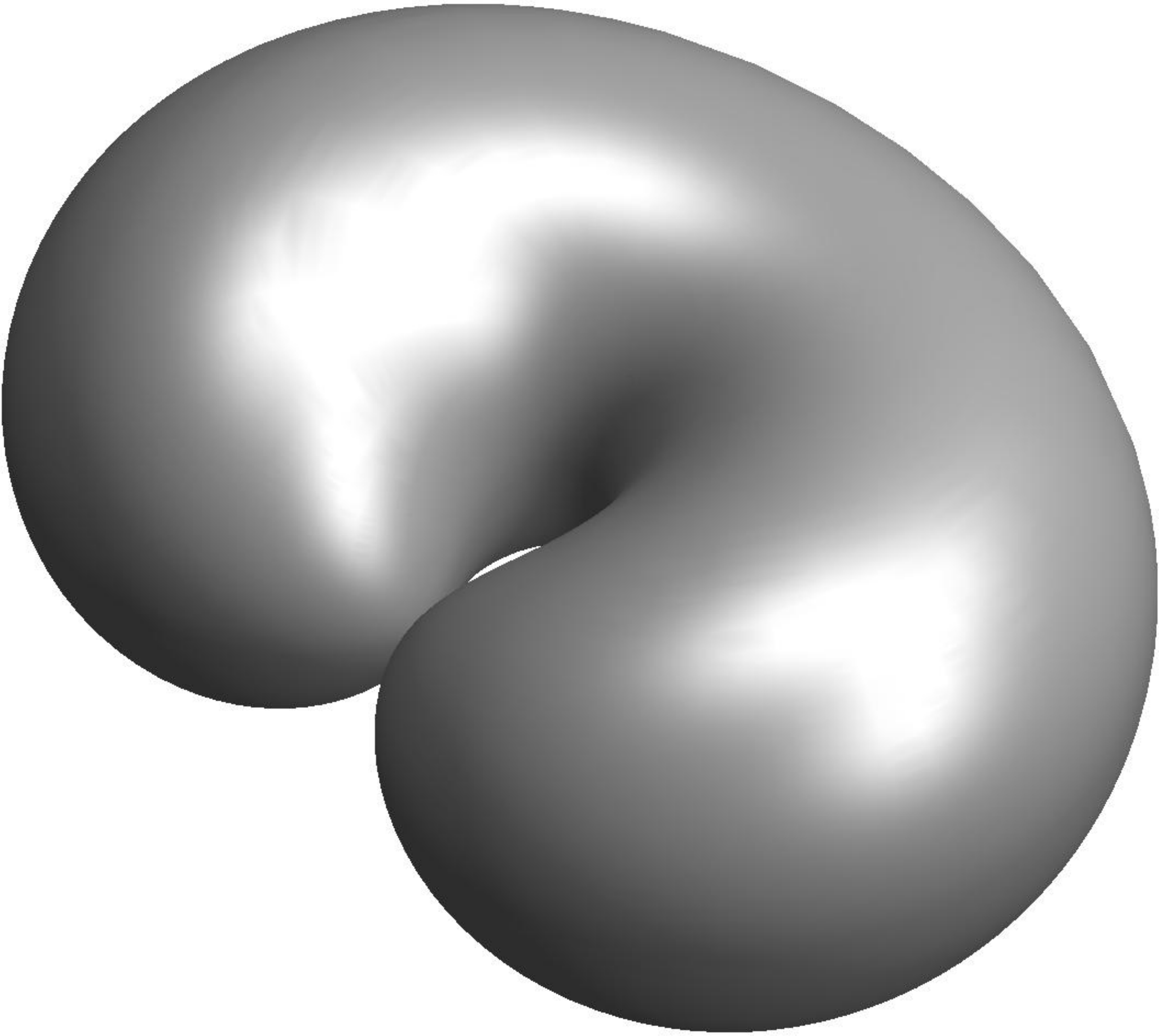}}
\caption{\small Surfaces obtained by inversion centered (a) inside the catenoid and (b) outside.}
\label{fig5}
\end{center}
\end{figure}

The symmetric defect geometry is characterized by a single length scale, $S$, the length of the
geodesic $\tilde\Gamma$ connecting the two neighboring points in contact. There is a direct
relationship between $S$ and the radius of inversion $R$. To establish this relationship  for small
$S$ ($\ll \sqrt{A_0}$), note that the line element on the catenoid is given by
\begin{equation} \label{linarelempolar}
ds^{2}=\rho^{2}(Z)\left(dZ^2 + d\varphi^2\right)\,;
\end{equation}
thus it is conformally flat, with conformal factor given by the polar radius
$\rho$. Its counterpart on the surface, $d\tilde s^2$, is given by
\begin{equation}
d\tilde s^2 = \frac{R^4}{\left(Z^2+P^2(\rho,\varphi)\right)^2} \, ds^2\,;
\end{equation}
thus the length of $\tilde \Gamma$ is given by
\begin{equation} \label{SR}
S = 2 R^2 \displaystyle\int\limits^\infty_{0} \frac{\rho}{Z^2 +
(\rho+1)^2} dZ \approx 1.206 R^2 \,.
\end{equation}
A defect geometry with a given value of $S$ is obtained from a unit catenoid by choosing $R$ appropriately.
For a given value of $S$ there is a unique symmetric geometry with two points held together.
\vskip1pc\noindent
If the center of inversion lies strictly on the catenoid, the area of
the inverted geometry will be infinite.  The normalization of the area would then require a
vanishing value of $R$ so the separation between the two points vanish, the geometry spherical.
It is worth considering a little more carefully this limit in order to interpret Eq. (\ref{SR}).
Let us, therefore, consider the inversion of the catenoid in a sphere centered on a point ${\bf x}_0 = x_0 \,
{\bf i}$ along the $X$ axis. The total area is given in terms of $x_0$ and $R$ by the expression,
\begin{equation} \label{Inversearea}
 A_0 = 2 R^4 I_A(x_0)\,, \qquad I_A(x_0) = \displaystyle{\int \limits_{Z=0}^{\infty}}
\displaystyle{\int \limits_{\phi=0}^{2 \pi}} \frac{\mathrm{d}Z \mathrm{d} \phi \, \rho^2}{\left(Z^2
+ \rho^2 -2 x_0 \cos \phi + x_0^2 \right)^2}\,.
\end{equation}
Normalizing the total area by the radius of the spherical vesicle, so $A_0=4\pi$ in Eq. (\ref{Inversearea}), implies the functional relationship between $R$ and $x_0$ illustrated in Fig. \ref{fig4}(a). $R$, of course, vanishes at $x_0 = 1$.
The corresponding distance between the poles along the membrane is given by
\begin{equation}
 S = 2 R^2 I_S(x_0)\,, \qquad I_S(x_0) = \displaystyle{\int \limits_{0}^\infty} \frac{\mathrm{d}Z \,
\rho}{Z^2 + (\rho + x_0)^2}\,.
\end{equation}
To an excellent approximation, $I_S(x_0)$ is given by $ I_S = 1.13073/(x_0 + 0.874017)$.
Equation (\ref{SR}) is reproduced when $|x_0|\ll 1$  which justifies the approximation.
For a fixed area,  $R$ is given as a function of $x_0$ by Fig. \ref{fig4}(a), which determines $S$
as a function of $x_0$ as illustrated in Fig. \ref{fig4}(b).
\vskip1pc \noindent
We confine ourselves to $|x_0|\ll 1$.  As $|x_0-1|$ is increased, the separation between the two
points on the vesicle increases. This delocalization of the defect is reflected in a progressively
less spherical geometry.  This is illustrated in Fig. \ref{fig5}, where the defect geometry is
represented for $x_0=0.5$ and $x_0=2$. In the former, the geometry approximates the axially
symmetric discocyte discussed in Refs. \cite{guvencastro, invcat};  in the latter it resembles a
sausage  with its two ends tied together.  As $x_0$ becomes large, however, these mathematical
curiosities are likely to become increasingly unreliable representations of physically realistic
defects. The details will be described elsewhere \cite{Trinoid}.
\vskip1pc\noindent
The connection between the value of $x_0$ and the reduced volume is given by
\begin{equation}
V_0 = 2 R^6 I_V(x_0)\,, \qquad I_V(x_0) = \frac{1}{3}\displaystyle{\int \limits_{Z=0}^{\infty}}
\displaystyle{\int \limits_{\phi=0}^{2 \pi}} \frac{\mathrm{d}Z \mathrm{d} \phi \, \rho^2 \left(1-
x_0 \cos \phi \,  \sech Z - Z \tanh Z \right)}{\left(Z^2 + \rho^2 -2 x_0 \cos \phi + x_0^2
\right)^3}\,.
\end{equation}
In Fig. \ref{fig4}(c) we plot the normalized volume $v_0 = V_0/V_S$ as a function of $x_0$, with
$V_S$ the volume of the unit sphere. This normalized volume is multivalued in an interval of the
normalized distance between the defects $s_0$, as shown in  Fig \ref{fig4}(d). In the regime we are
interested in, with $|x_0-1|\ll1$, the delocalization  of the defect correlates directly with the
deflation of the vesicle. There is a a unique symmetric two-finger defect, as well as a pinched
counterpart for each value of $V_0$. For larger values of $x_0$ this duality between fingers and
pinches breaks down.  In particular, the maximally deflate vesicle bearing two fingers is a
sausage-like geometry, not the two touching spheres that occur in the limit
$x_0\to\infty$.\footnote{Intriguingly, there is a narrow band of values of $s_0$ in which a volume
three fold degeneracy exists, with two distinct two-finger geometries and a single pinch; see Fig.
\ref{fig4}(d).} The pinched counterpart terminates at $x_0=0$ in the symmetric discocyte.

\subsection{Curvatures} \label{curvatures}

\begin{figure}[htb]
\begin{center}
\subfigure[Catenoid]{\includegraphics[scale=0.13]{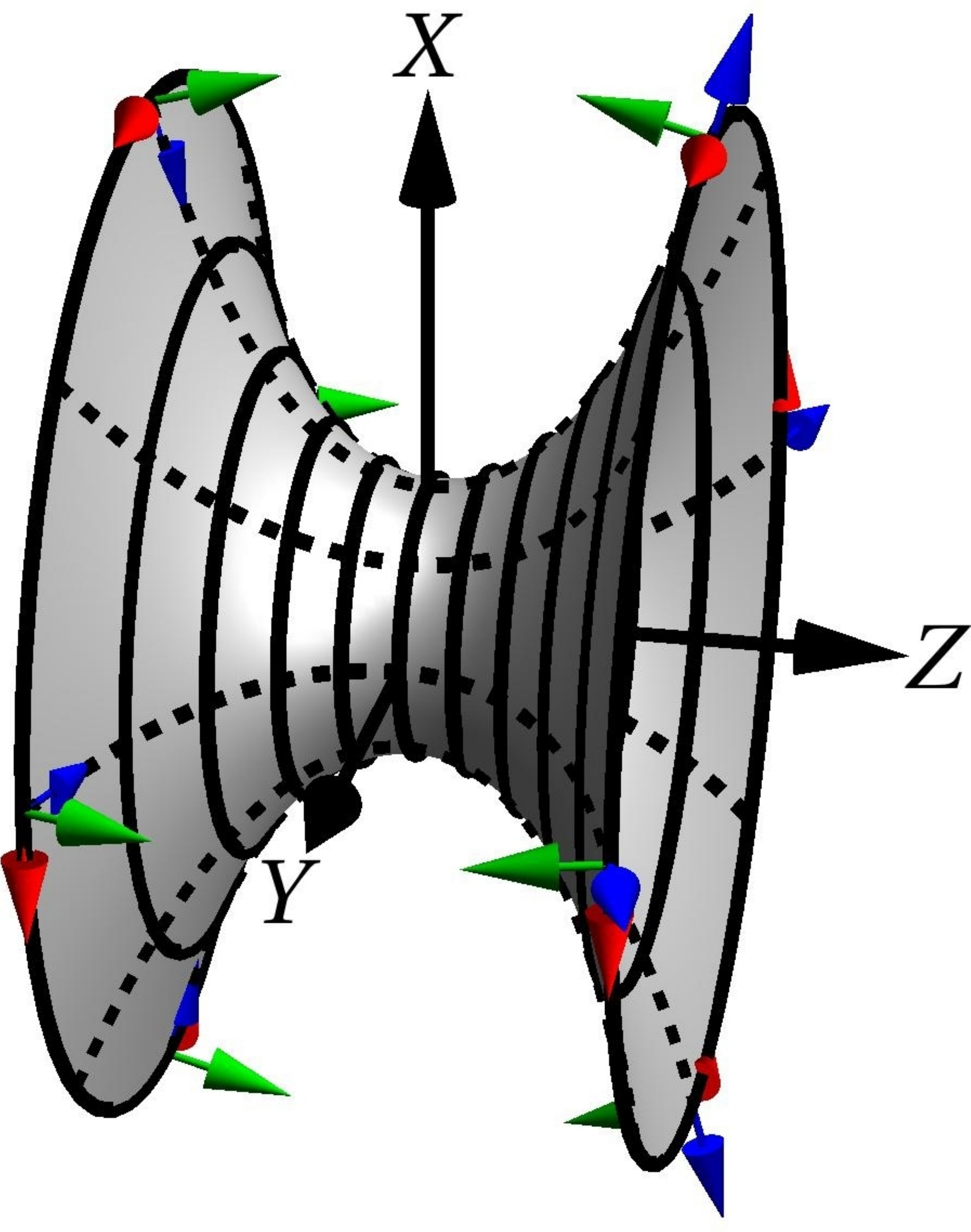}} \hfil
\subfigure[Defect top view]{\includegraphics[scale=0.125]{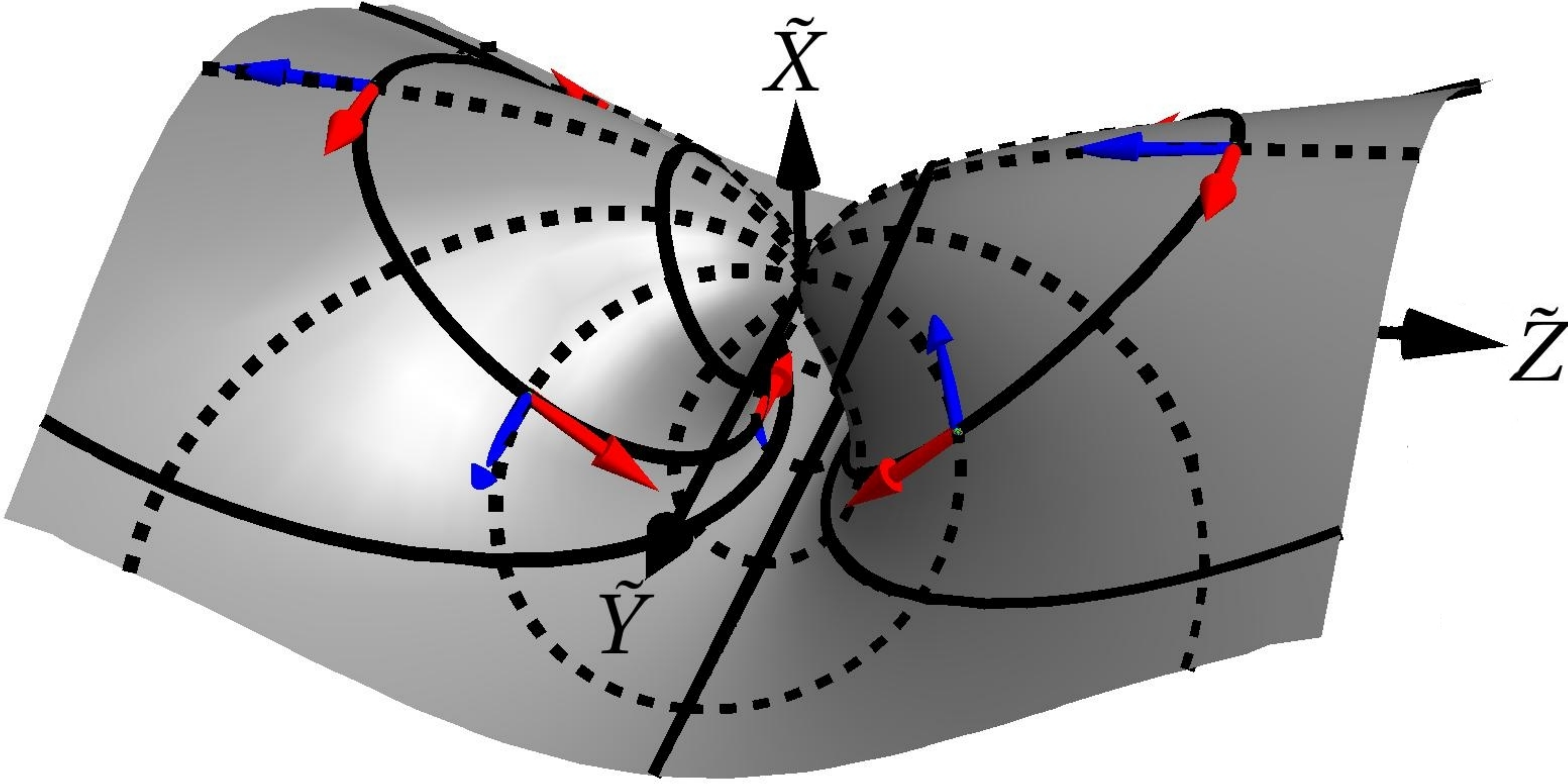}}
\hfil \subfigure[Defect bottom
view]{\includegraphics[scale=0.125]{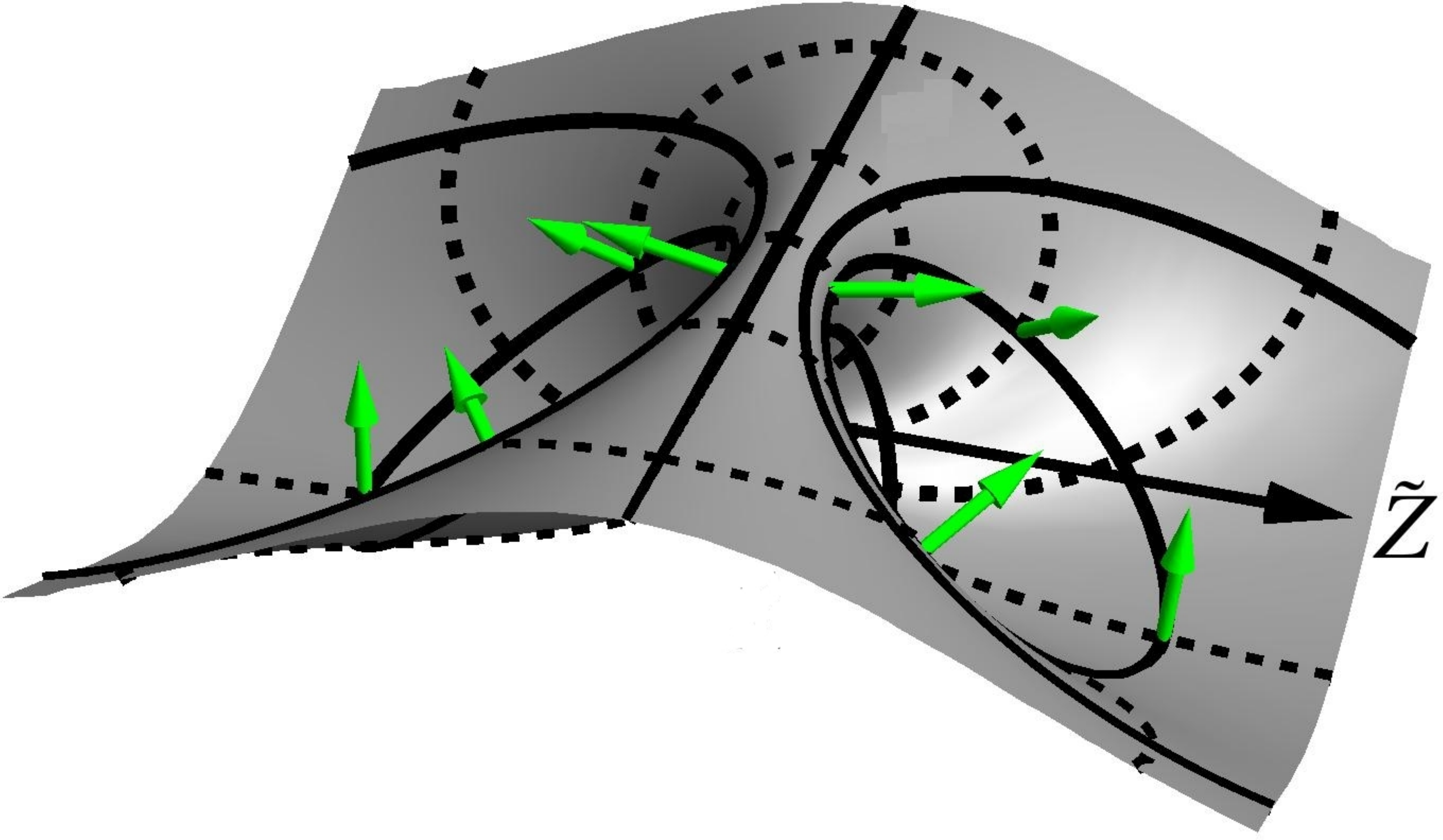}} \caption{\small (Color online)
Principal curves and directions on the catenoid and on the defect. (a) On the catenoid, the
meridians are catenaries (dashed lines) with tangent vector ${\bf V}_\perp$ represented by blue
(dark) arrows; the parallels (solid circles) with the tangent vector ${\bf
V}_\parallel$ are represented by red (medium gray) arrows (they are oriented
anticlockwise with respect to the symmetry axis). The normal vectors are represented by green
(light gray) arrows. (b) Top and (c) bottom views of the principal curves on the
defect. The catenaries get mapped to closed curves which make self-contacts at the poles,
represented by dashed lines, whereas the parallel circles transform to rotated circles surrounding
the poles, indicated by solid lines. Red (medium gray), blue (dark), and green (light gray) arrows represent the mapping of the
principal directions and the normal vector onto the defect.}
\label{fig6}
\end{center}
\end{figure}

The curvature of the defect will play a role not only in determining its energy but also in
determining the stresses. Singularities in the curvature signal external forces. The curvature is
characterized by two scalars, its extremal values, as well as the directions--mutually
orthogonal--along which they occur. The intersections of the defect surface with the mirror planes
are both directions of curvature as well as geodesic.
\vskip1pc \noindent
On the catenoid, the maximum curvature $C_\parallel$ is achieved along the parallel directions
(circles of constant $Z$ in Fig. \ref{fig6}(a), and the minimum curvature $C_\perp$ along meridians
(catenaries in Fig. \ref{fig6}(a)). They are given by $C_\parallel =1/\rho^2= -C_\perp$, confirming
that the catenoid is a minimal surface, i.e. $K = C_\parallel + C_\perp =0$. Its Gaussian curvature
${\cal K}_G = C_\parallel C_\perp$ is given by
\begin{equation} \label{gaussMS}
{\cal K}_G = - 1/\rho^4\,.
\end{equation}
The principal directions (along the parallel circles and along the meridians) align with the adapted basis
vectors ${\bf e}_\varphi=\partial_\varphi {\bf X}$ and ${\bf e}_Z = \partial_Z {\bf X}$,
\begin{subequations}
\begin{equation} \label{knoidsprindirec}
{\bf V}_\parallel = \left(-\sin \varphi, \cos \varphi,0\right) \,; \qquad
{\bf V}_\perp = 1/\rho \left(\sinh Z \cos \varphi, \sinh Z \sin \varphi
,1\right)\,.
\end{equation}
\end{subequations}
In \ref{appinvgeom}  it is shown that the counterparts on the defect of these two
vectors fields are given by an appropriate reflection:
\begin{equation} \label{tildeV}
\widetilde{\bf V}_{i}= {\sf R} \, {\bf V}_{i}\,, \qquad i= \parallel, \;
\perp
\end{equation}
where ${\sf R}= \mathbbm{1} - 2 \hat{\bf X} \otimes \hat{\bf X}$ represents a
reflection in the plane passing through the origin and orthogonal to ${\bf
X}$, and  $\hat{\bf X}$ denotes the corresponding unit vector.
\vskip1pc\noindent
The integral curves of these two vector fields on the defect are represented in Figs. \ref{fig6}(b)
and \ref{fig6}(c). In particular, observe that the parallel circles on the catenoid get mapped to
two nested families of circles, each of which encloses a single pole; these circles degenerate into
the straight line $\widetilde\Gamma_\pi$ on the mirror plane, $\tilde Z=0$.  The meridians get
mapped to a single family of nested closed curves, each of which passes through the point $O$ where
self contact is made.
\vskip1pc \noindent
Three regions displaying features of particular interest can be identified on the defect:
the neighborhood of the poles, the ridge and valley geometry connecting the poles, and the asymptotic
region where it approximates the large sphere. Although the global defect geometry does not lend
itself to a Monge parametrization in terms of a unique height function above a plane, it is possible
to describe each of these regions separately in terms of a height function above an appropriate reference
plane.

\subsection{Height function representations} \label{Height}

\subsubsection{Near region} \label{near}

The region $\widetilde Z \ll 1$ originates in the compactification of the exponential growing ends
of the catenoid with $Z\gg1$ under inversion in a sphere centered on the point ${\bf i}$. This
region on the catenoid can be approximated by the height function $h_1=\ln 2 \rho$ ($\rho\gg1$)
above the $X-Y$ plane, ($X=\rho \cos \varphi$ and $Y=\rho \sin \varphi$). Its counterpart on the
defect under inversion in a sphere of radius $R$ then has the height function representation
\begin{equation} \label{MongeXpole}
\widetilde h_1 = - R^2 \, \widetilde \rho^2 \ln \left(
\frac{\widetilde \rho}{2} \right)\,, \qquad \widetilde \rho = \frac{1}{\rho}\,,
\quad \widetilde
\varphi = \varphi\,,
\end{equation}
above the plane $\widetilde Z=0$ ( $\widetilde X = R^2 \widetilde \rho \cos \widetilde \phi$ and
$\widetilde Y = R^2 \widetilde \rho \sin \widetilde \phi$).
The two curvatures diverge logarithmically at the origin as $\widetilde \rho \rightarrow 0$
\begin{equation}
\label{Clog}
\widetilde{C}_{{\parallel}\,,\perp} \approx -\frac{2}{R^2} \left(\ln
\frac{\widetilde{\rho}}{2} +1\pm \frac{1}{2} \right)\,.
\end{equation}
However their difference remains finite
\begin{equation}
\widetilde{C}_{\parallel}-\widetilde{C}_{\perp} \approx -\frac{2}{R^2} \,.
\end{equation}

\subsubsection{Ridge/Valley region}

The ridge and valley between the poles originates in the neighborhood of the point $-{\bf i}$ on the
catenoid diametrically opposite the center of inversion. The latter is described by the
saddle-shaped height function above the plane $X=-1$: $h_2=- r^2/2 \, \cos 2 \phi$, where
$r=\sqrt{Y^2+Z^2}$, and $\phi = \arctan Y/Z$. Its inversion in a sphere of radius $R$ centered at
${\bf i}$ is a parabolic cylinder described by a height function above the plane $\widetilde
X=-R^2/2$, ($\widetilde Y = R^2/4 \, \widetilde r \sin \widetilde \phi$, $\widetilde Z = R^2/4 \,
\widetilde r \cos \widetilde \phi$)
\begin{equation}
\widetilde h_2 = \frac{R^2}{4} \, \widetilde r^2 \cos^2 \widetilde \phi \,,
\quad \widetilde r= r\,, \quad \widetilde
\phi = \phi\,,
\end{equation}
or $\widetilde h_2 = 4 \widetilde Z^2 /R^2$.

\subsubsection{Far region} \label{asymptgeom}

\begin{figure}[htb]
\begin{center}
\includegraphics[scale=0.125,origin=c]{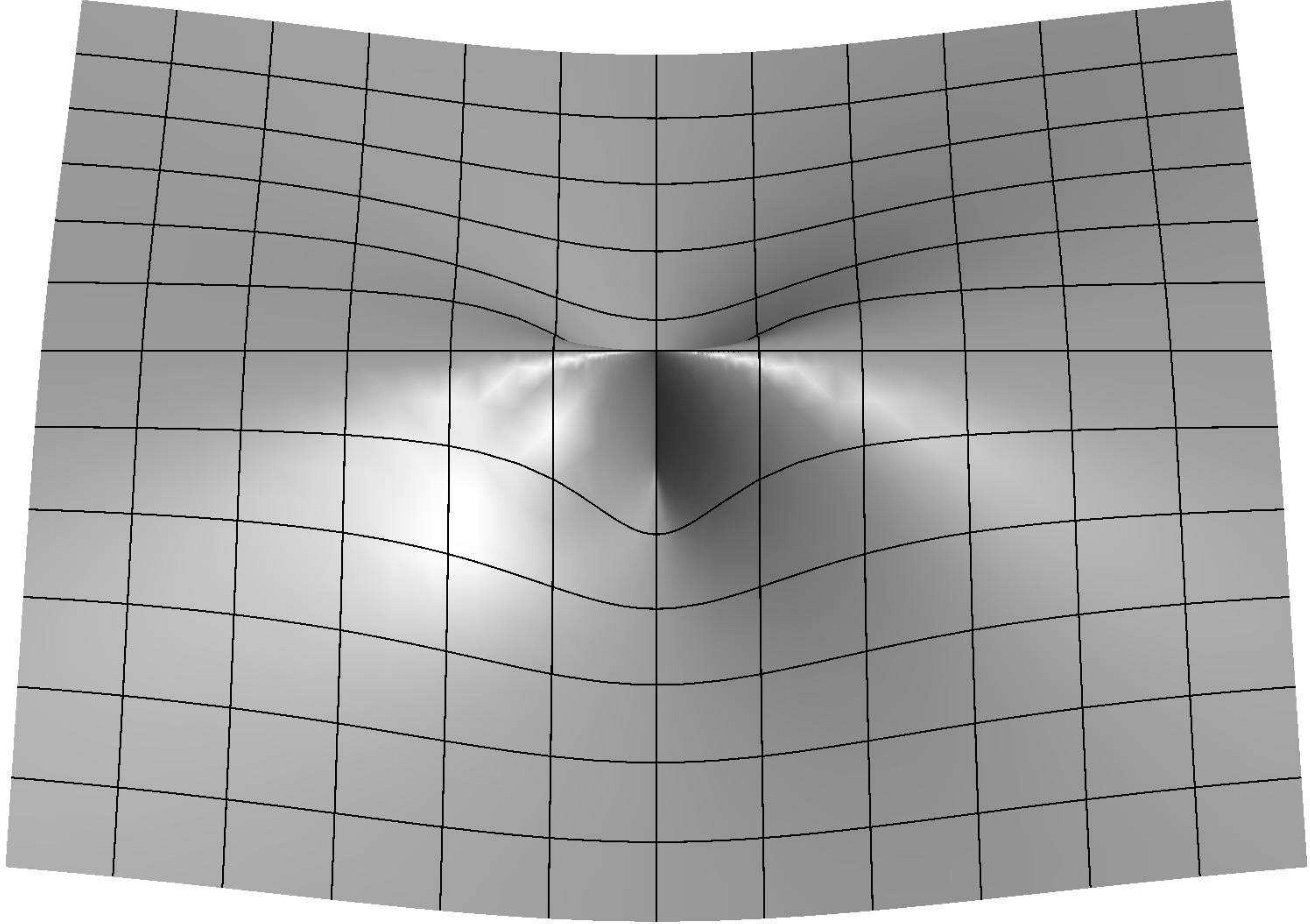}
\caption{\small The surface given by the biharmonic height function
$\widetilde h= 1/2 \cos 2 \widetilde \phi$ provides an accurate description of the defect  in the
far region.} \label{fig7}
\end{center}
\end{figure}

The infinite region remote from the origin originates in the neighborhood of the center of
inversion ${\bf i}$ on the catenoid. The latter is described by the height function $h_3 = r^2/2
\cos 2 \phi$. Inversion of this region in a sphere of radius $R$ centered at ${\bf i}$ can also be
described by the height function above the the asymptotic
plane $\widetilde{X}=0$ ($\widetilde Y = R^2 \widetilde r \sin \widetilde \phi $ and $\widetilde Z
= R^2 \widetilde r \cos \widetilde \phi $)
\begin{equation} \label{MongeYasym}
 \widetilde h_3= \frac{R^2}{2} \cos 2  \widetilde \phi\,, \quad \widetilde r=
\frac{1}{r} \,, \quad \widetilde \phi = \phi\,,
\end{equation}
which is independent of the radial distance $\widetilde r$ on the base plane. Thus in the
asymptotic region, with $\widetilde r \gg 1$, the defect interpolates between the two orthogonal
lines, $\widetilde{X}=R^2/2$, $\tilde Y=0$ and $\widetilde{X}=-R^2/2, \widetilde{Z}=0$;  its height
set by the scale $S$ (see Eq. (\ref{SR})).  This geometry is illustrated in  Fig. \ref{fig7}. Unlike
the exact defect, the geometry is symmetrical (modulo a rotation by $\pi/2$) with respect to the
plane $\widetilde X=0$.
In the small gradient approximation, the bending energy is proportional to
\begin{equation}
H \approx \frac{1}{2}\int d{\bf x}\, (\nabla^2 h)^2\,,
\end{equation}
where $d{\bf x}$ is the element of area on the plane and $\nabla^2$ is the corresponding Laplacian.
This is minimized when $h$ is biharmonic on the plane, or $(\nabla^2)^2  h =0$. Note that all
three height functions are biharmonic on their respective planes, consistent with the fact that each
of the three surface geometries locally minimizes bending energy. The catenoid from which
they stem is a minimal surface minimizing locally the surface area.\footnote{\sf  The surface area
excess is given in the small gradient approximation by
\begin{equation}
A \approx \frac{1}{2}\int d{\bf x}\, (\nabla h)^2\,,
\end{equation}
so minimal surfaces satisfy the Laplace equation locally. }
\vskip1pc \noindent
In cylindrical coordinates, $h = \cos 2 \phi$, is the only biharmonic function
independent of $r$. It belongs, with $k=2$, to the family of biharmonic functions, $h=\cos k
\phi/r^{k-2}$ with a $k$-fold dihedral symmetry, obtained by the inversion in a sphere centered at
the origin of the harmonic height functions $h^{(k)}=r^k \cos k \phi$, $k\geq 2$, representing
$k$-th order saddles (a monkey saddle is represented by $k=3$).

\section{Defect Energy} \label{sectenergy}

\begin{figure} [htb]
\begin{center}
\subfigure[$\tanh \widetilde
K$]{\includegraphics[scale=0.12]{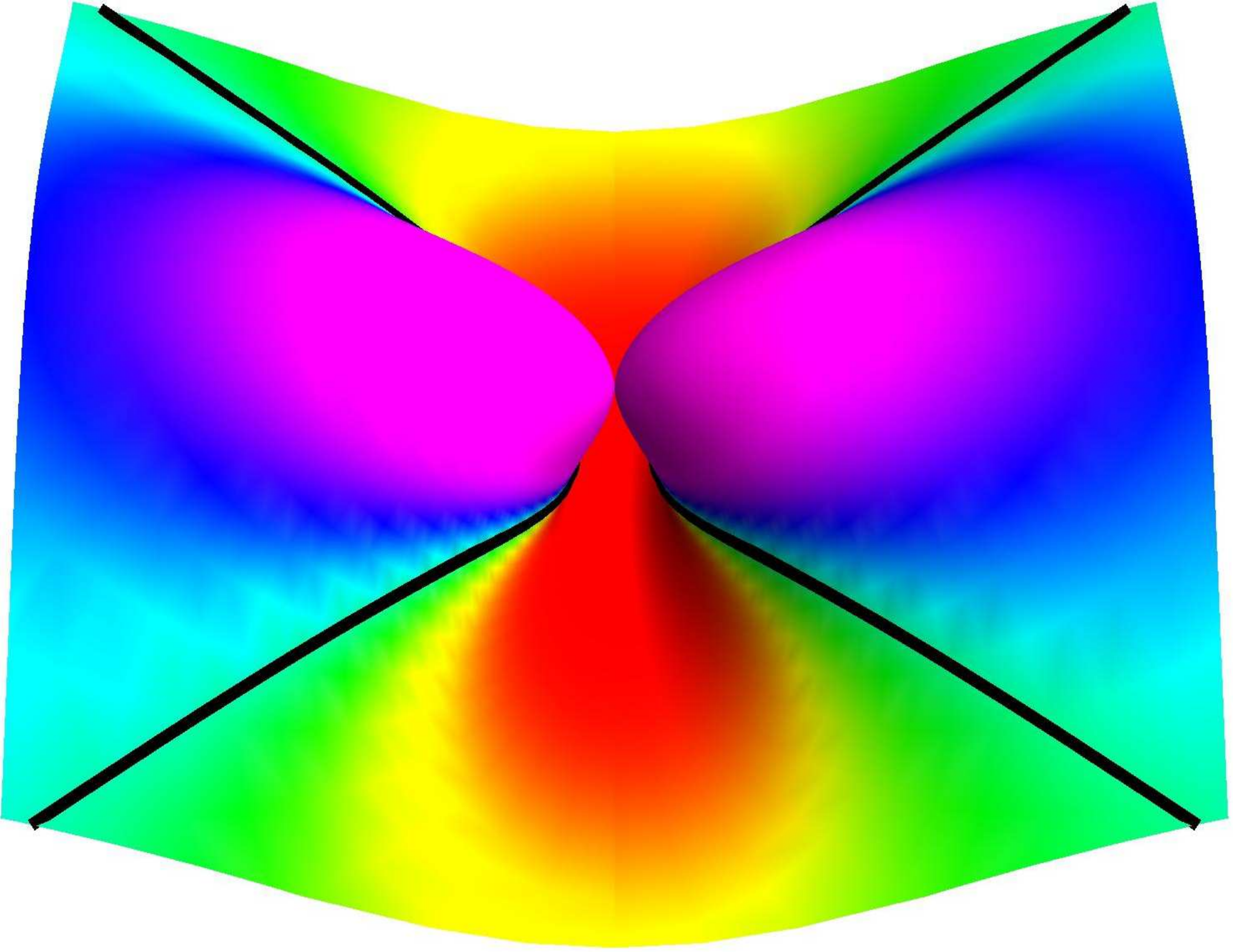}} \hfill
\subfigure[$\tanh 1/2 \widetilde
K^2$]{\includegraphics[scale=0.12]{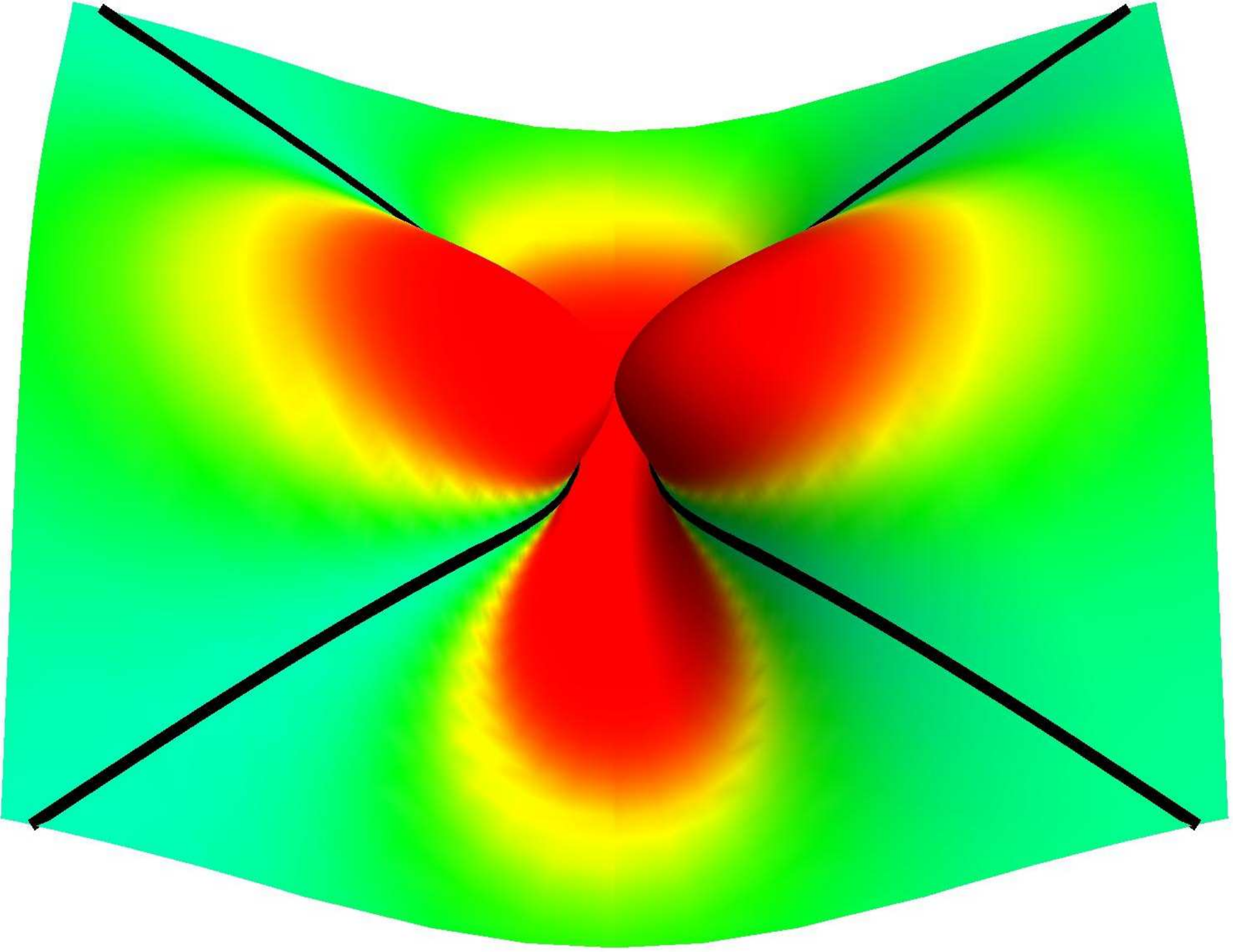}} \hfill
\subfigure[$\tanh \widetilde{\cal K}_{G}$]{\includegraphics[scale=0.12]{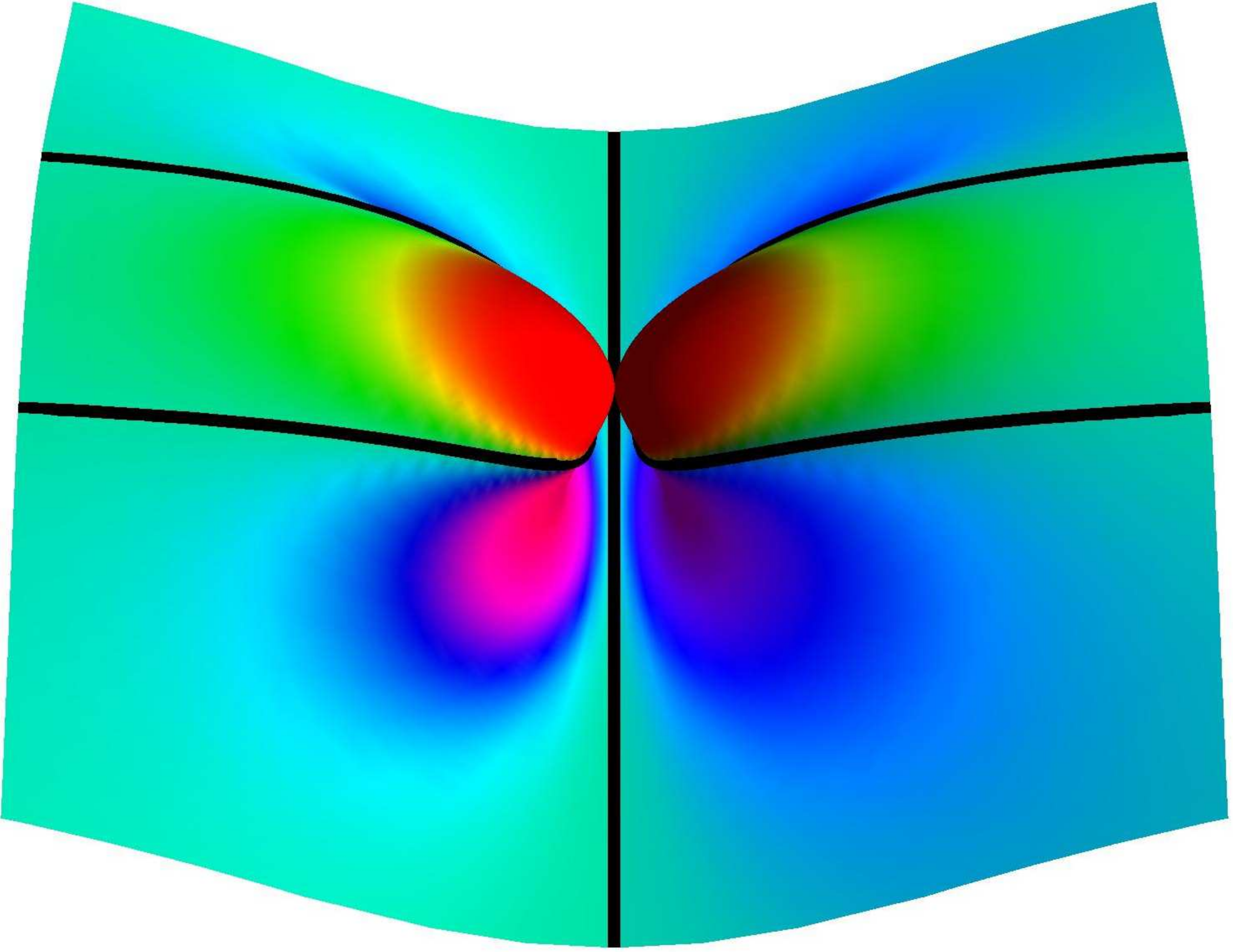}}\\
\includegraphics[scale=0.15]{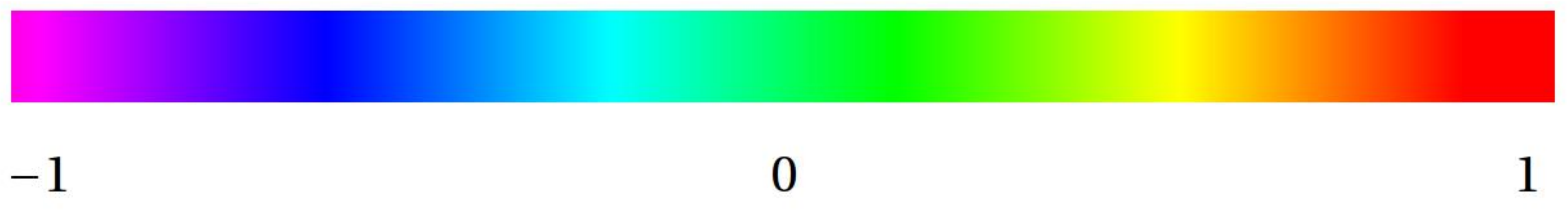}
\caption{\small (Color online) (a) $\widetilde K$, (b) $\widetilde K^2/2$, and (c)
$\widetilde K_G$. These functions vanish along black lines.
} \label{fig8}
\end{center}
\end{figure}

While the bending energy is not itself invariant under inversion, it {\it is} modulo a
contribution proportional to the topological term, $H_{GB}$. This is because $H_1$ can be cast as a
linear combination of $H_{GB}$ and the manifestly conformally invariant energy $H_W= 1/2\int dA\,
(C_\parallel - C_\perp)^2$. As a consequence, equilibrium states are mapped to equilibrium states
under inversion. The defect geometry constructed in Sec. \ref{sectinvknoids} is thus also an
equilibrium state.
\vskip1pc \noindent
It is simple to show that the bending energy of the inverted surface is given by (see also Ref. \cite{Trinoid})
\begin{equation} \label{bendeninvsurf}
\widetilde{H} = \kappa ( H_1 + 2 (\widetilde{H}_{GB} - H_{GB}))
+\bar\kappa\, \widetilde{H}_{GB} \,.
\end{equation}
Whereas $H_1$ vanishes for a catenoid it does not on its inverted counterpart. Thus the total
bending energy of a surface obtained by inversion in a sphere of a minimal surface is proportional
to the difference of two topological contributions.
\vskip1pc \noindent
In general the topology will change under inversion if the surface is infinite or the point of inversion lies
on the surface. The catenoid has the topology of a punctured plane with Gauss-Bonnet energy
$H_{GB}= -4 \pi$; the vesicle with the defect has a spherical topology,
with $\widetilde{H}_{GB} = 4 \pi$ (see \ref{appGB}). Thus the bending energy of the defect is given
by $\widetilde{H} = 4 \pi (4 \kappa + \bar{\kappa})$, depending on both physical
parameters.
\vskip1pc \noindent
In particular, $\widetilde H$ is independent of $S$. Although the total energy is constant, the
distribution of energy in the membrane
\begin{equation} \label{Hinv}
\widetilde {\cal H}=  \frac{1}{2}\kappa \, \widetilde K^2 + \bar\kappa \,
\widetilde K_G\,,
\end{equation}
is concentrated in the vicinity of the defect center; it also depends on $\bar\kappa$.
Note first that the mean curvature of the defect is proportional to the support function on the
catenoid (see \ref{appinvgeom})
\begin{equation} \label{Kinv}
 \widetilde K = \frac{4}{R^2} {\bf X} \cdot {\bf n}\,, \qquad {\bf X} \cdot {\bf
n} = 1-\cos \varphi \, \sech Z- Z \tanh Z\,.
\end{equation}
Its (scaled) value, as well as $\widetilde K^2/2$, are illustrated in Figs. \ref{fig8}(a) and
\ref{fig8}(b). It vanishes along the curves $\cos \varphi = \cosh Z -Z \tanh Z$, indicated in black
in Fig. \ref{fig8}(a) and \ref{fig8}(b).
\vskip1pc \noindent
Likewise, the Gaussian curvature on the defect is given by
\begin{equation} \label{KGinv}
 \widetilde K_G = \frac{1}{R^4} \left[4 \left({\bf X} \cdot {\bf
n}\right)^2-\left(Z^2+P(\rho,\phi)^2\right)^2 \sech^4 Z \right]\,.
\end{equation}
$\widetilde K_G$ is positive  in a neighborhood of the two poles that extends on its dorsal side
to infinity, a region one would expect to be under tension; it vanishes where $4 \cos \varphi =
\sech Z \left(Z^2+1\right) + 3 \cosh Z- 2 Z \sinh Z$,  and along the axis of symmetry $Z=0$,
indicated in black in Fig. \ref{fig8}(c). In Sec. \ref{near} it was seen that the two principal
curvatures diverge logarithmically near the poles (see Eq. (\ref{Clog})), so the bending
energies do also: $\widetilde K^2, \widetilde K_G \approx \ln^2  (\widetilde \rho/2)/
R^4$. The corresponding densities $\sqrt{g} K^2$ and $\sqrt{g} K_G$, however, not
only remain finite but vanish at the poles. Asymptotically, $\widetilde {\cal H}$ decays as
$1/\tilde r^4$, as is easily confirmed using Eq. (\ref{MongeYasym}).

\subsection{Stability and the zero mode} \label{slippage}

A feature of the conformal symmetry of this problem is that defect states are energetically stable.
Indeed, this is implied by the stability of a catenoid described by the energy $H_B[{\bf X}] $. The
second variation of the energy is given by \cite{Defo}
\begin{equation}
\delta^2 H_B = \kappa \, \int dA\, \Phi {\cal L} \Phi\,,
\end{equation}
where--for a minimal surface--the self-adjoint operator ${\cal L}$ is given by ${\cal L}= (- \nabla^2 + 2 K_G)^2$. $\Phi$ is the normal displacement. ${\cal L}$ is manifestly positive, which implies the stability of the catenoid as a fluid membrane.  Conformal  invariance now implies equality, term by term, in perturbation theory between
the energy of the minimal surface and that of its counterparts obtained by inversion in a sphere. They are, thus, also
guaranteed to be energetically stable.
\vskip1pc\noindent
This is not the end of the story however. For the energy is degenerate.  Modulo
Euclidean motions, there will be a one-parameter family of defect states of a given area and volume with the
same energy. Distinct geometries with a fixed area  are labeled by the position of the center of
inversion, $\mathbf{x}_0$ on the quadrant, $x_0,y_0\ge 0$;  fixing the volume constrains $x_0$ in
terms of $y_0$.
The construction of symmetric states, with $y_0=0$, was discussed in Sec. \ref{character}. When $y_0 \ne 0$  the mirror symmetry between the poles is broken.
\vskip1pc\noindent
This degeneracy implies the existence of a zero or Goldstone mode satisfying ${\cal L} \Phi=0$.  It can be constructed explicitly as a linear combination of  a rescaling and a special conformal transformation, $\Phi={\bf n}\cdot \delta {\bf X}$, where
$\delta {\bf X}$ is given by Eq. (\ref{sct}).\footnote {In this section, ${\bf X}$ refers to the defect geometry.}
It is clear that modulo Euclidean motions, the axial symmetry of the catenoid implies that $\mathbf{c}$ has two independent components corresponding to a displacement of the center of inversion on the quadrant.
The first order constraints on the area and the volume, $\int dA\, K\,\Phi=0$ and $\int dA\, \Phi=0$, impose two constraints, leaving a single degree of freedom. On a mirror symmetric state this mode generates displacement along ${\bf j}$ which breaks the mirror symmetry.
\vskip1pc\noindent
This degeneracy implies that the two points tied together still possess a freedom to migrate
across the surface, at no energy cost, while preserving the area and volume. However the  volume
constraint keeps them apart.  If it is relaxed, there will be a second zero mode corresponding to
displacements of $\mathbf{x}_0$ along the direction $\mathbf{i}$ . As we described in Sec.
\ref{character} this corresponds to changing the geodesic distance $S$ between the poles;
in particular, there will now be a path between the defect state and a sphere, generated by
displacing the center of inversion towards the surface of the catenoid. This same mode allows
the membrane to slip out of the defect. As we will discuss in Sec. \ref{externalF}, however,
this limit is not smooth, involving a divergence in the force between the two poles. Before it is
reached one would expect the defect to collapse.

\section{Stresses and transmitted forces on the defect}

The stress tensor on the membrane is given by ${\bf f}^a = {\sf f}^{ab}\,{\bf e}_b + f^a \, {\bf
n}$ \cite{Stress, Auxiliary} (see also Refs. \cite{Jenkins, Steigman, Lomholt}), where (in units of
$\kappa$)
\begin{equation} \label{eq:stressdef}
{\sf f}^{ab} = K \left(K^{ab} - \frac{1}{ 2} \, K\,g^{ab}\right) \,, \quad f^a = -\nabla^a K\,,
\end{equation}
where ${\bf e}_a$, $a=1,2$  are the two tangent vectors adapted to the parametrization, ${\bf n}$ is
the normal vector, and $g_{ab}$ and $K_{ab}$ are the the metric and extrinsic curvature tensors,
defined in \ref{appinvgeom}. Under a local deformation of the equilibrium surface, ${\bf X}\to {\bf X}+
\delta {\bf X}$, its energy changes by an amount
\begin{equation} \label{eq:stressintuit}
\delta H= - \int dA\, {\bf f}^a \cdot \nabla_a \delta {\bf X}\,.
\end{equation}
This stress depends only on the local geometry. There is no local stress associated with the
Gauss-Bonnet bending energy.  The tangential stress is quadratic in $K_{ab}$ and thus assumes its
maximum and minimum  values along the directions of curvature.  In a source free region, ${\bf f}^a$
is conserved in equilibrium so \cite{Stress}
\begin{equation} \label{eq:conslawfa}
\nabla_a \,{\bf f}^a =0\,,
\end{equation}
where $\nabla_a$ is the covariant derivative compatible with the metric tensor.
\vskip1pc \noindent
Consider a curve $\gamma$ partitioning the surface in two regions, such as one of those illustrated in
Fig. \ref{fig9}(a). The force per unit-length on one side of this curve (darker region in Fig.
\ref{fig9}), exerted by the other side (lighter region in Fig. \ref{fig9})
is given in terms of the stress tensor by ${\bf f}_\perp = -l_a {\bf f}^a$, where
$l_a=g_{ab} l^b$ are the covariant components of the vector ${\bf l}$, the outward normal to
$\gamma$ tangent to the surface, {\it i.e.}, directed towards the region exerting the
force.\footnote{If ${\bf t}$ is the unit tangent vector and ${\bf n}$ the surface normal along the
loop, then ${\bf l} = {\bf t} \times {\bf n}$ and $\{{\bf t},{\bf n},{\bf l}\}$ form a right-handed
trihedron (the Darboux frame). Since ${\bf t}$ and ${\bf l}$ are tangent vectors to $\Sigma$ they
can be expanded with respect to the adapted basis vectors, $\{{\bf e}_a\}$ as ${\bf t}=t^{a}{\bf
e}_{a}$ and ${\bf l} = l^{a}\bf e_a$.} If $\gamma$ is closed, the
total force is given by the line integral \cite{MDG}
\begin{equation}
\label{eq:totforcelineint}
 {\bf F} = \oint_{\gamma} ds\, {\bf f}_\perp \,,
\end{equation}
where $s$ is the arc-length along $\gamma$. If $\gamma$ is contractible to a point or if it does not
enclose sources of stress, this integral vanishes.

\subsection{Stress distribution in the defect: quadratic approximation} \label{stressquad}

Before attempting to describe the fully non-linear distribution of stress in the defect, it is
useful to, first, provide an approximate description of this distribution in each of the three regions
described in Sec. \ref{Height} in terms of a height function. \footnote {To unburden the
notation, set $R=1$ and rotate the geometry so the base plane coincides with the plane $Z=0$
parametrized by the radial and azimuthal coordinates $r=\sqrt{X^2+Y^2}$ and $\phi = \arctan
(Y/X)$.} It is simple to check that gradients remain small in the regions of interest, i.e. $|\nabla
h| \ll 1$ where $\nabla$ represents the gradient on the base plane, so a quadratic
approximation to the stress tensor in terms of gradients and Hessians of the height function is
legitimate.
\vskip1pc \noindent
In the quadratic approximation, the tangent and normal vectors to the surface can be decomposed with
respect to an orthonormal basis adopted to a Cartesian description of the appropriate base plane $\{
\hat{\bf E}_x$,$\hat{\bf E}_y$, ${\bf k}\}$:  ${\bf e}_i=\hat{\bf E}_i+\partial_i h\, {\bf k}$ and
${\bf n}=-\partial_i h\, \hat{\bf E}_i+{\bf k}$\footnote{The normal points toward positive $Z$}. The
metric tensor is given by $g_{ij}=\delta_{ij}+\nabla_i h\, \nabla_j h$. In this approximation, the
extrinsic curvature tensors is given by minus the Hessian of the height function $K_{ij}= -\nabla_i
\nabla_j h$.
\vskip1pc \noindent
The quadratic approximation to the stress tensor (\ref{eq:stressdef}) is now given by
${\bf f}^i = T^{ij} \hat{\bf E}_j + N^i {\bf k}$ (compare with Ref. \cite{Fournier}),
where
\begin{equation} \label{Tijdef}
T^{ij}= \nabla^2 h \, \left(\nabla^i \nabla^j h - \frac{1}{2} (\nabla^2
h)\delta^{ij} \right)- \nabla^i (\nabla^2 h) \, \nabla^j  h\,, \qquad N^i =
\nabla^i \nabla^2 h\,.
\end{equation}
Both $T^{ij}$ and $N^i$ are conserved, or $\nabla_i T^{ij}=0$ and
$\nabla_i N^i=0$, when the height function is biharmonic.
\footnote{Note that $T_{ij}$, unlike the genuine
tangential stress ${\sf f}^{ab}$, is not symmetric. Consequently, $\nabla_i T_{ji}\ne 0$. In
addition, its trace $T^i_{\phantom{i}i}$ is different from zero, whereas $f^a_{\phantom{a}a}$
vanishes.} These equations are equivalent to the conservation law (\ref{eq:conslawfa}) in the
quadratic approximation.
\vskip1pc \noindent
Let ${\bf T}$ and ${\bf L}$ represent the unit vectors along the projections of the vectors ${\bf
t}$ and ${\bf l}$ (introduced below Eq. (\ref{eq:totforcelineint})) onto the base plane, so they
form an orthonormal basis adapted to the curve $\Gamma$ obtained by projection of the surface curve
$\gamma$ onto this plane.  Equation (\ref{eq:totforcelineint}) then implies that the force per unit
length transmitted across the curve $\gamma$ on the surface is given, in this approximation, by
${\bf f}_\perp = - L_i {\bf f}^i = - L_{i} T^{i j} {\bf E}_j - L_i N^i {\bf k}$.

\subsubsection{Poles: $h_1 = -r^2 \ln (r/2)$, $r \ll 1$} \label{poleforce}

Note that $\nabla^2 h_1 = -4 (\ln (r/2)+1)$, and $(\nabla^2)^2  h_1= -8 \pi \delta (r)$. Thus,
near the poles, the stress tensor  on the base plane has components
\begin{equation}
 T^{rr} = -4 \ln \frac{r}{2} \,, \quad  T^{r \phi} = T^{\phi r} =0 \,, \quad
T^{\phi \phi} = - \frac{4}{r^2} \ln \frac{r}{2}\,;\\
\end{equation}
with counterparts normal to the plane
\begin{equation}
N^r=-\frac{4}{r} \,,  \quad N^\phi =0\,.
\end{equation}
Construct a circle of constant $r_c$ on the base plane, oriented such that  ${\bf T}= \hat{\bf
E}_{\phi}= (-\sin \phi,\cos \phi,0)$ and ${\bf L} = -\hat{\bf E}_{r}= (-\cos \phi,-\sin
\phi,0)$\footnote{This gives the force exerted by the interior region with $r<r_c$, containing the
pole, on the exterior region with $r>r_c$}. By projecting the stress tensor onto ${\bf L}$, one
identifies the force transmitted per unit length across this circle to be given by
\begin{equation}
\label{fperpr}
{\bf f}_{\perp r}= -4 \left(\ln \frac{r_c}{2}  \hat{\bf E}_r + \frac{1}{r_c}\, {\bf k}\right)\,.
\end{equation}
Thus, the external radial force, viewed from the base plane,  can be decomposed into a sum of radial
compression and a vertical force associated with a source pulling (down) on the pole at its center.
This is a striking illustration of the  shortcomings of the Monge representation: One would expect
radial tension rather than compression.  It should be remembered, however, that $\hat{\bf E}_r$ is
not the tangent to the surface,
nor is ${\bf k}$ its normal. To resolve this discrepancy, it is necessary to expand the radial force
with respect to a basis of tangent vectors adapted to the surface: ${\bf e}_r = \hat{\bf E}_r-2 r
\ln (r/2) {\bf k}$ and ${\bf n} = 2 r \ln (r/2) \hat{\bf E}_r +{\bf k}$. One then recasts Eq.
(\ref{fperpr}) as
\begin{equation}
{\bf f}_{\perp r} = 4 \left(\ln \frac{r_c}{2}  {\bf e}_r - \frac{1}{r_c}\, {\bf n}\right)\,,
\end{equation}
which indicates that the outer region is subjected to radial tension plus a normal force pulling on
the pole. This coincides with the result obtained below in the non-linear analysis, Sec.
\ref{stresinvsurf}.
\vskip1pc \noindent
If ${\bf f}_{\perp r}$ is integrated along the circle one identifies the total external force acting
on the pole to be $-8 \pi \hat{\bf k}$.  In fact, this quadratic analysis provides the correct exact
non-linear result (\ref{eq:Finvfinal}) derived below. This is because the quadratic approximation
becomes exact at the pole.
\vskip1pc \noindent
Now consider a radial line on the base plane along direction $\hat{\bf E}_r$ and with outward
normal ${\bf L} = \hat{\bf E}_\phi$\footnote{The force is that  exerted from a region with a value
of $\phi$ below a given value on that with a higher value.}. The force per unit length transmitted
across this line is given by
\begin{equation}
{\bf f}_{\| \phi}=  4 \ln \frac{r}{2} \, \hat{\bf E}_\phi\,,
\end{equation}
which implies compression along the azimuthal direction.
\vskip1pc \noindent
Near $r=0$,  the azimuthal compression is equal in magnitude to the radial tension, as measured
along the surface. This result hints at a more general result: Tension in one direction is always
accompanied by compression in an orthogonal direction. As discussed in greater detail in Sec.
\ref{stresinvsurf}, this is a direct consequence of the vanishing trace of the tangential stress,
i.e. ${\sf f}^{a}_{\phantom{a}a}=0$. Surprisingly, this is not evident if tangents are replaced by
their projections on the plane, where one registers compression in both directions, radial and
azimuthal.  As was seen, this apparent discrepancy  is an artifact of the Monge parametrization;
specifically the third term in the definition of $T^{ij}$, Eq. (\ref{Tijdef}),  originating in the
projection of the normal stress onto the base plane, contributes a trace to $T^{ij}$: $T_{ii} = -
\nabla_i (\nabla^2 h) \, \nabla_i  h$.  Furthermore, $\nabla_i (\nabla^2 h) \, \nabla_j  h$  diverges
near the pole as $1/r$.  Thus, whereas the distinction between ${\bf n}$ and ${\bf k}$ disappears
near the pole, where the gradients of $h$ are small (${\bf n}-{\bf k} \approx 2 r\ln r/2 \,{\bf
e}_r) $),  $T^{ij} ({\bf n}-{\bf k})$ diverges logarithmically.  This example illustrates how the
Monge representation can yield counterintuitive results whenever curvature singularities  are
involved. One needs to be careful to distinguish between tangent vectors and their projections onto
the plane. This kind of gauge artifact is sidestepped in the covariant non-linear analysis presented
in Sec. \ref{stresinvsurf}.

\subsubsection{Valley: $h_2 = 1/4\, X^2$}

The tangential stress is diagonal with constant components
\begin{equation}
T_{XX} =  1/8\,, \quad T_{XY} = T_{YX} = 0\,, \quad T_{YY} =  -1/8\,.
\end{equation}
Furthermore, because the Laplacian is constant, $\nabla^2 h_2=1/2$, both normal components vanish,
$N_X = N_Y =  0$. The force transmitted across the valley (constant $X$, along direction $\hat{\bf
E}_x$ with ${\bf L} = \hat{\bf E}_x$) is given by ${\bf f}_{\perp X} = -1/8 \hat{\bf E}_x$, so it is under compression. Likewise the force along the valley (constant $Y$ along direction $\hat{\bf E}_y$, with ${\bf L}=\hat{\bf E}_y$) is ${\bf f}_{\| Y}= 1/8 \hat{\bf E}_y$ so it is under tension along its length.

\subsubsection{Asymptotic region:  $h_3 = 1/2 \cos 2 \phi$}
\label{asymptension}
In the asymptotic region, the components of the stress tensor on the plane are ($\nabla^2 h_3 = -2/r^2 \cos 2 \phi$)
\begin{equation}
T^{rr} = -\frac{2}{r^4} \cos^2 2 \phi \,, \quad T^{r \phi} = - T^{\phi r}
=\frac{1}{r^5} \sin 4 \phi \,, \quad T^{\phi \phi} =  \frac{2}{r^6} (1+\sin^2 2
\phi)\,.\\
\end{equation}
$T^{ij}$ is not diagonal. This reflects the fact that the principal directions do not
coincide asymptotically with the radial and azimuthal directions. The corresponding components normal
to the plane are
\begin{equation}
N^r = \frac{4}{r^3} \, \cos 2 \phi \,,  \quad N^\phi = \frac{4}{r^4} \sin 2
\phi\,.
\end{equation}
Projecting onto the azimuthal and radial directions, one finds the forces per unit length
transmitted across curves with constant $r$ (${\bf L}=\hat{\bf E}_r$) and $\phi$ (${\bf L}=\hat{\bf
E}_\phi$) are given respectively by \footnote {\sf The presence of a component along $\hat{\bf
E}_\phi$ in ${\bf f}_{\perp r}$, as well as component along $\hat{\bf E}_r$ in ${\bf f}_{\| \phi}$
reflects the fact that the vectors $\hat{\bf E}_r, \hat{\bf E}_\phi$ and ${\bf k}$ do not coincide
asymptotically with the physically significant vectors, $\widetilde{\bf V}_\perp, \widetilde{\bf
V}_\|$, and $\widetilde{\bf n}$, described in Sec. \ref{curvatures}.}
\begin{subequations}
\begin{eqnarray}
{\bf f}_{\perp r}  &=& -\frac{1}{r^4}\,\sin \, 4 \phi \, \hat{\bf E}_\phi + \frac{2}{r^4}\, \cos^2
\, 2 \phi \, \hat{\bf E}_r - \frac{4}{r^3} \, \cos \, 2 \phi \, {\bf k}\,, \label{frasympt}\\
{\bf f}_{\| \phi} &=& - \frac{2}{r^4}\,(1+\sin^2 \, 2 \phi) \hat{\bf E}_\phi +
\frac{1}{r^4} \sin \, 4 \phi\, \hat{\bf E}_r - \frac{4}{r^3} \, \sin \, 2 \phi \, {\bf k}\,.
\label{fphiasympt}
\end{eqnarray}
\end{subequations}
The tangential forces decay as $r^{-4}$, their normal counterparts as $r^{-3}$, confirming the
localization of the stress in the neighborhood of the poles.

\subsection{Distribution of stress: exact results} \label{stresinvsurf}

The distribution of stress in the defect is completely encoded in the catenoid geometry.
It is straightforward  to show, using results presented in \ref{appinvgeom} that (see also Ref. \cite{Trinoid})
\begin{equation} \label{eq:finvgen}
\widetilde{\bf f}^a = \left(\frac{|{\bf X}|}{R}\right)^6
\left({\sf R \, \bf f}^a + \frac{4}{|{\bf X}|^2}
\left(K^{ab}- \frac{1}{2} K g^{ab}\right)\, {\bf w}_{b}\right)\,,
\end{equation}
where \cite{ConfJ, Laplace}
\begin{equation} \label{eq:f0surf}
{\bf w}_{a} = {\bf X} \times \, ({\bf n} \times {\bf e}_a) \,.
\end{equation}
Because ${\bf w}_{a}$ is orthogonal to ${\bf X}$, ${\sf R} \,{\bf w}_{a}= {\bf
w}_{a}$. Whereas the stress tensor ${\bf f}^a$ vanishes in a minimal surface
with $K=0$,  it will not vanish in its inverted counterpart:
\begin{equation} \label{eq:finv}
\widetilde{{\bf f}}^a = 4\frac{|{\bf X}|^4}{R^6} \, K^{ab}\,{\bf w}_{b}\,.
\end{equation}
It can also be shown (see Ref. \cite{Trinoid}) that if ${\bf f}^a$ is
conserved then $\widetilde{\bf f}^a $ is also,  except at the points in contact
at $\tilde{\bf X}=0$, represented  by $Z\rightarrow \pm
\infty$. It is evident that the stress is concentrated in the neighborhood of
these points. Its divergence indicates the presence of sources on the right hand
side of Eq. (\ref{eq:conslawfa}). These sources are identified with   the
external forces enforcing the contact constraint. In the absence of these
forces the equilibrium would collapse to a sphere with a vanishing stress.

\subsubsection{External forces and torques on the poles}\label{externalF}

\begin{figure}[htb]
\begin{center}
\subfigure[
$\widetilde{V}_\parallel$]{\includegraphics[scale=0.12]{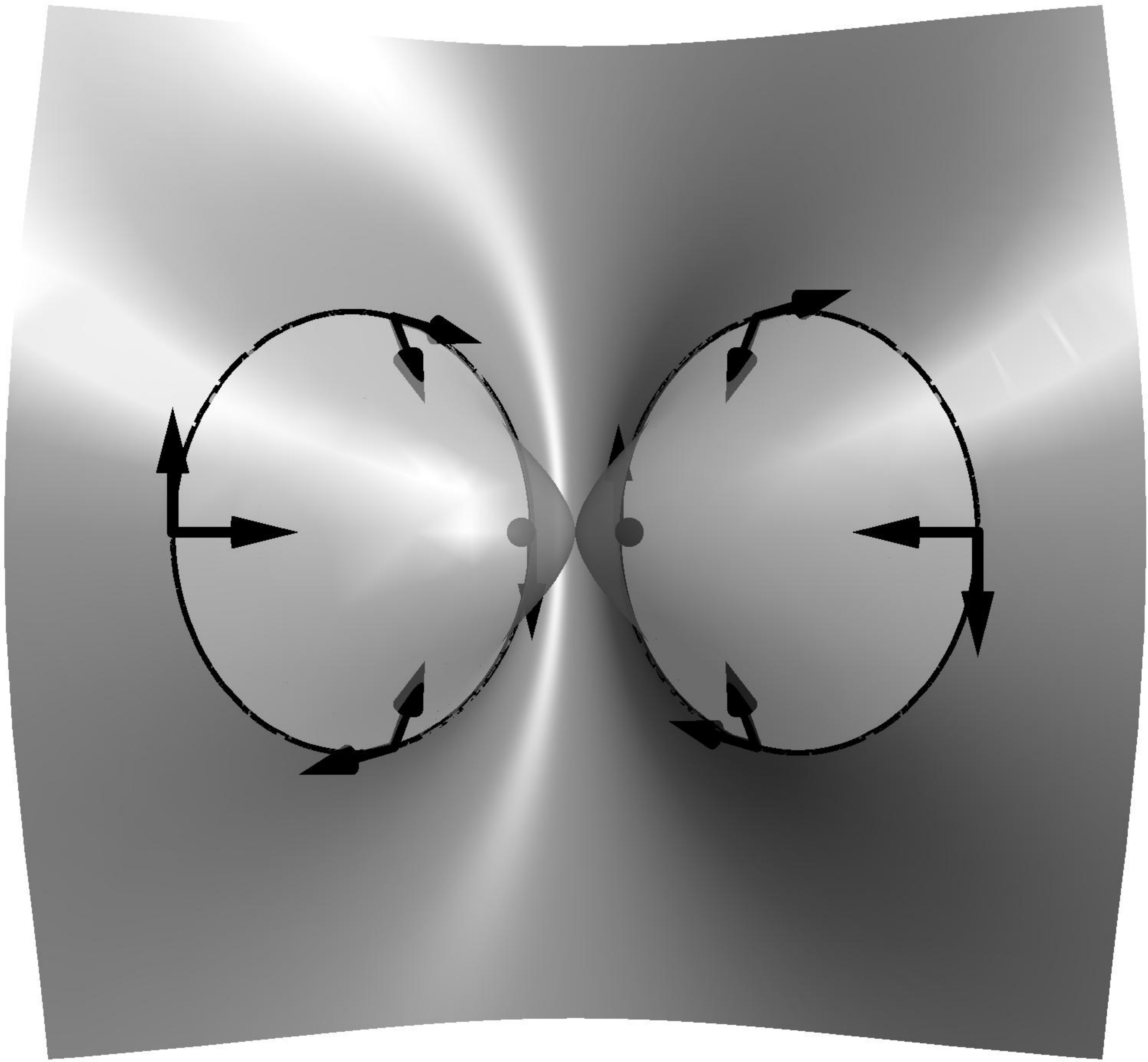}}
\qquad \qquad
\subfigure[$\widetilde{V}_\perp$]{\includegraphics[scale=0.12]{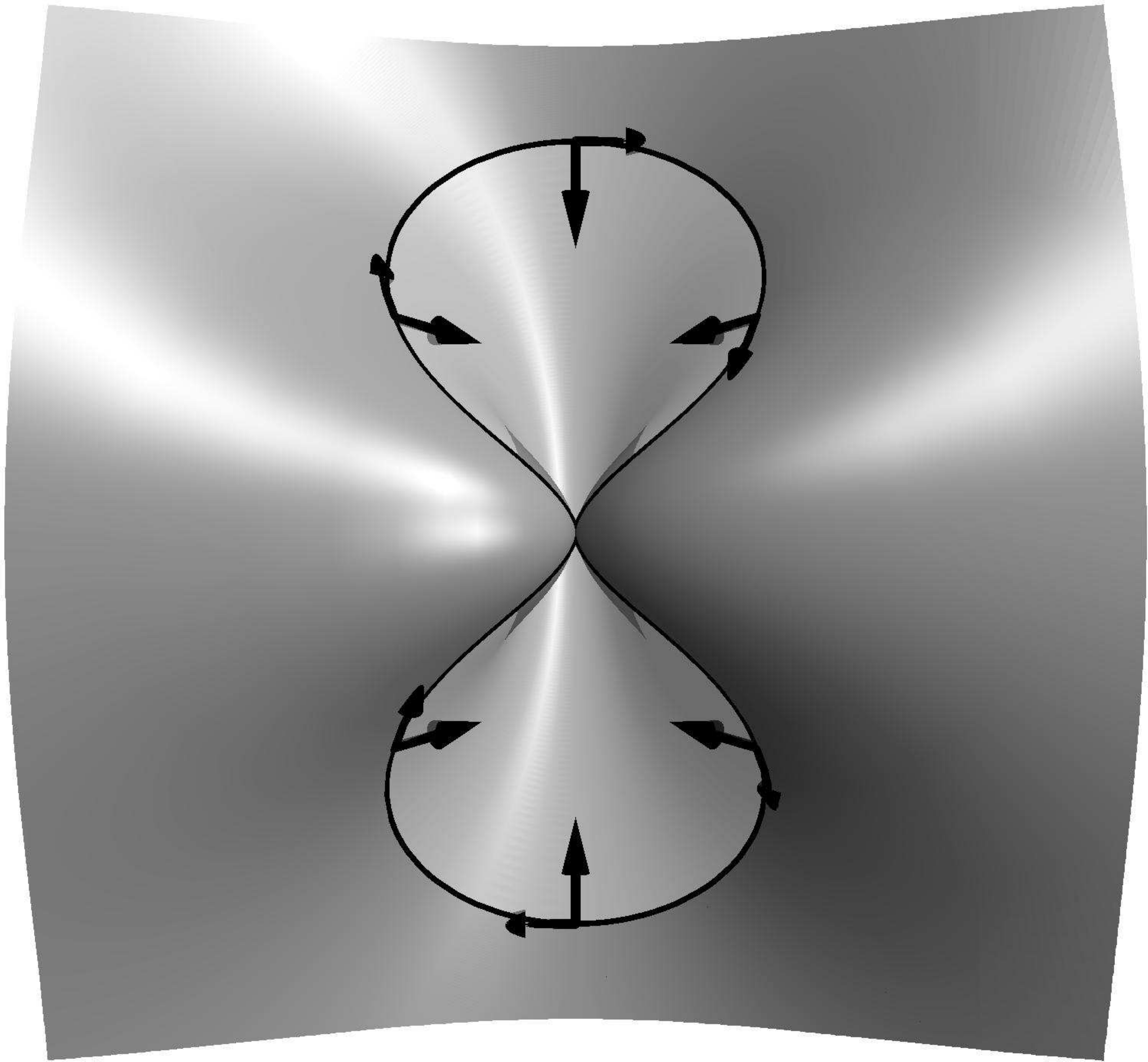}}
\caption{\small  Closed integral curves of $\widetilde{V}_\parallel$ and $\widetilde{V}_\perp$.
} \label{fig9}
\end{center}
\end{figure}

The external force pulling the poles together is determined by evaluating the integral of the
stress Eq. (\ref{eq:totforcelineint}) along a loop enclosing each point \cite{MDG}.\footnote{The
forces supporting the defect cannot be resolved by examining the asymptotic physics. A contour
completely encircling the defect will register a vanishing total force.} An appropriate contour is
one of the closed circular integral curves of $\widetilde{V}_\parallel$ (indicated using red
red(medium gray) arrows in Fig.\ref{fig6}(b)), illustrated in Fig. \ref{fig9}(a).\footnote{The appropriately oriented
boundary has the Darboux basis $\widetilde{\bf t}= \sign \widetilde Z \, \widetilde{\bf
V}_\parallel$ and $\widetilde{\bf l}= \sign \widetilde Z \,\widetilde{\bf V}_\perp$.}
The mirror symmetry in the plane $Z=0$ can now be exploited to deform this circular contour into the
straight line $\widetilde \Gamma_\parallel$ running along $Z=0$ illustrated on Figs. \ref{fig3}(b)
and \ref{fig6}(b). The contribution to the force from the contour at infinity vanishes  due
to the fast asymptotic decay of the stress tensor (see Eq. (\ref{frasympt})). The external force on
the right hand pole ($Z>0$) is now given by\footnote{The outward normal to $\widetilde
\Gamma_\parallel$ traversed from $\infty$ ($\varphi=0$) to $-\infty$ ($\varphi=2 \pi$) is
$\widetilde{\bf l} = {\bf k}$. In particular the components of the conormal vector expressed in the
tangent basis are $\widetilde{l}^Z=|{\bf X}|^2/R^2$, $|{\bf X}|^2=2(1-\cos \varphi)$ and
$\widetilde l^\varphi=0$, so the only component of its associated covector $\widetilde {\bf l}$
is $\widetilde l_Z=1/\widetilde l^Z$. Also the line-element along $\widetilde \Gamma_\parallel$ is
$\widetilde{d s}=R^2/|{\bf X}|^2\, d\varphi$ and the only relevant component of the stress tensor is
given by $\widetilde{\bf f}^Z = 2 |{\bf X}|^6/R^6 \, {\bf k}$.}
\begin{equation} \label{eq:Finvfinal}
\widetilde{\bf F} = -\int d\widetilde{s} \; \widetilde{l}_Z \, \widetilde{\bf f}^Z
= -\frac{4}{R^2} \, {\bf k} \displaystyle\int\limits^{2 \pi}_0 d\varphi \left(1-\cos \varphi
\right)= -\frac{8 \pi}{R^2}\,{\bf k}\,.
\end{equation}
The detailed distribution of stress need not be known to determine this
force.  This is just as well, for the detailed distribution of stress is not simple.
This result also validates the analysis of the linearized theory
in Sec. \ref{poleforce} where a circular contour, collapsing to the point of application of the
external force,  was used to determine this force.
\vskip1pc \noindent
Note that there is no net vertical displacement of the vesicle. This is consistent with the fact
that the forces acting on the membrane are horizontal. Raising the vesicle with respect to the
``asymptotic plane'' $\tilde X=0$ would require the introduction of vertical external forces which would
need to be balance by counter forces.
\vskip1pc \noindent
Using Eq. (\ref{SR}) which relates the inversion radius $R$ to the length $S$ between the poles,
one obtains $\widetilde{F} S=9.6539 \pi$.  This force diverges
as the points approach each other along the surface in an inflated sphere.
Below some critical separation, the maximum normal tension supported by the membrane or by
the external agent holding the two points together will be exceeded and the defect will disintegrate, decaying into
a spherical membrane as its additional bending energy is dissipated.  The volume constraint
keeps the points apart on the surface ensuring the stability of the defect. If this constraint is relaxed, however, one would also expect a defect to possess a finite lifetime. This lifetime will, however,  depend on physical
properties of fluid  membranes that fall outside the scope of the geometric description provided in
this paper.
\vskip1pc \noindent
The torque tensor on a surface is given by ${\bf m}^a = {\bf X} \times {\bf f}^a +
K {\bf e}^a \times {\bf n}$ \cite{Stress}. Using the same conventions as before  the torque per unit length
exerted by one region onto another across $\gamma$ is ${\bf m}_\perp = -l_a {\bf m}^a$ and the
total torque $M = \int ds {\bf m}_\perp $. For the surface resulting from the inversion of a minimal
surface the torque tensor reads
\begin{equation}
 \widetilde{\bf m}^a = 4  \frac{|{\bf X}|^2}{R^4} \left( K^{ab} {\bf X} \times {\bf w}_b +
{\bf X} \cdot {\bf n} \, {\sf R} ({\bf e}^a \times {\bf n})\right)\,.
\end{equation}
Evaluating the total torque along the contour used to determine the force we find that
\begin{equation}
 \widetilde{\bf M} = -4 \displaystyle{\int\limits_0^{2\pi}} \frac{\mathrm{d}\varphi }{|{\bf X}|^2}
\left(C_\perp\, {\bf X} \times ( {\bf X} \times {\bf t}) -{\bf X} \cdot {\bf n} \, {\sf R} {\bf
t}\right)\, = {\bf 0}\,.
\end{equation}
There is no net torque acting at the poles. Although the intermediate states
interpolating between the initial state and the pinched membrane will be subject to torques, the
final state is free of them.

\subsubsection{Stress distribution revisited}

\begin{figure}[htb]
\begin{center}
\subfigure[$\widetilde{\sf f}_{\perp
\perp}$]{\includegraphics[scale=0.12]{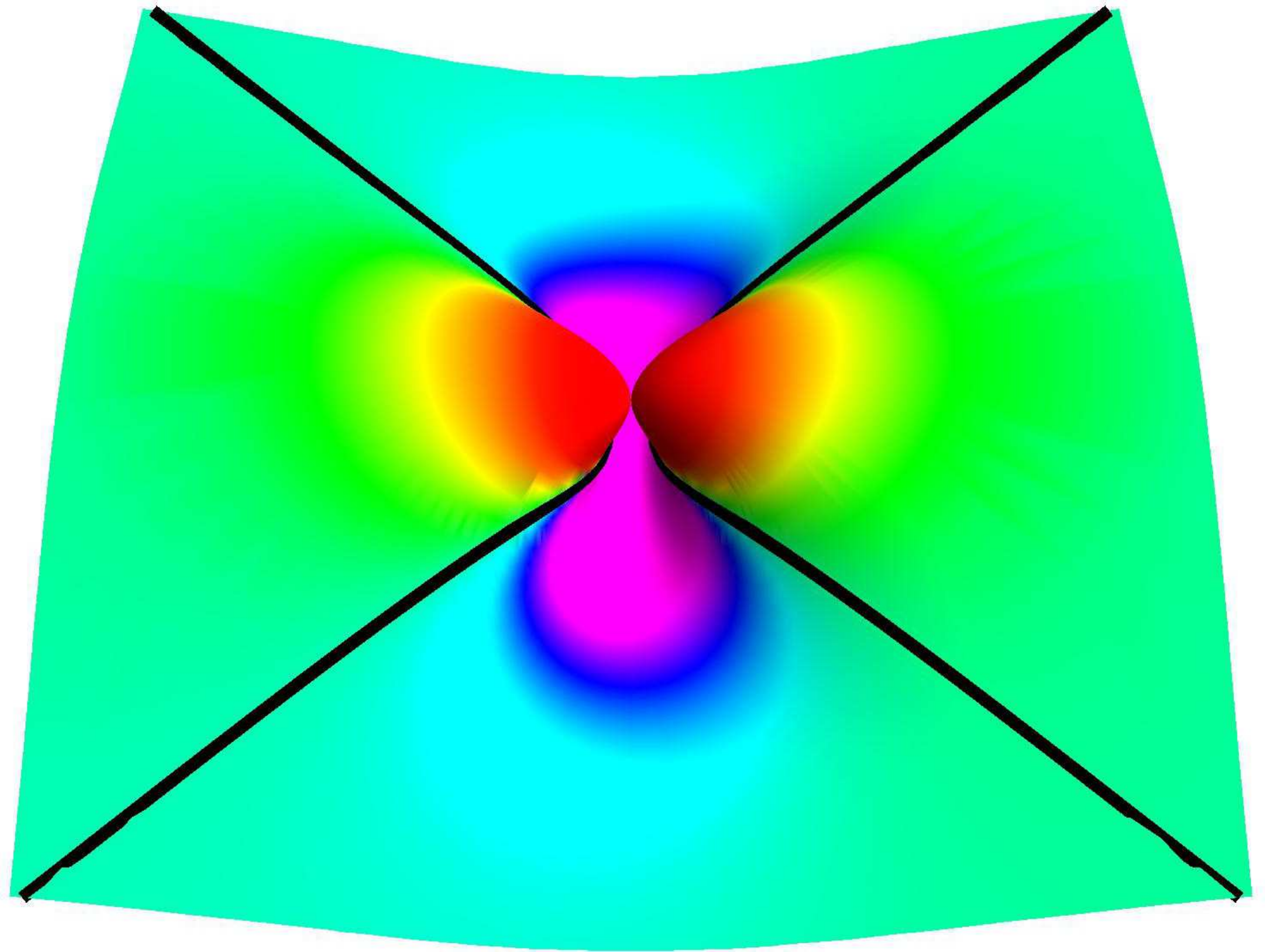}}
\hfil
\subfigure[$\widetilde{f}_\perp$]{\includegraphics[scale=0.12]{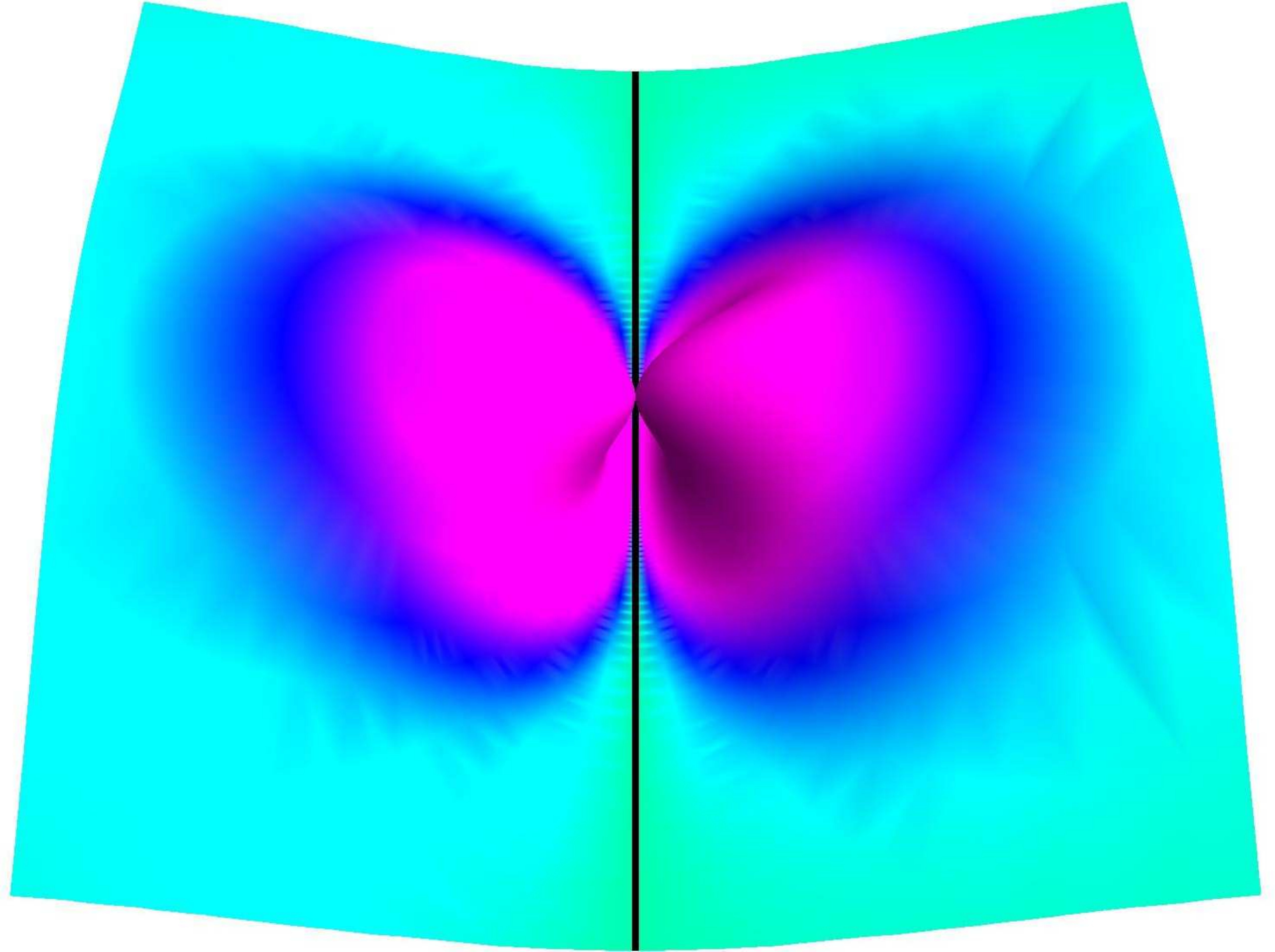}}
\hfil
\subfigure[$\widetilde{f}_\parallel$]{\includegraphics[scale=0.12]{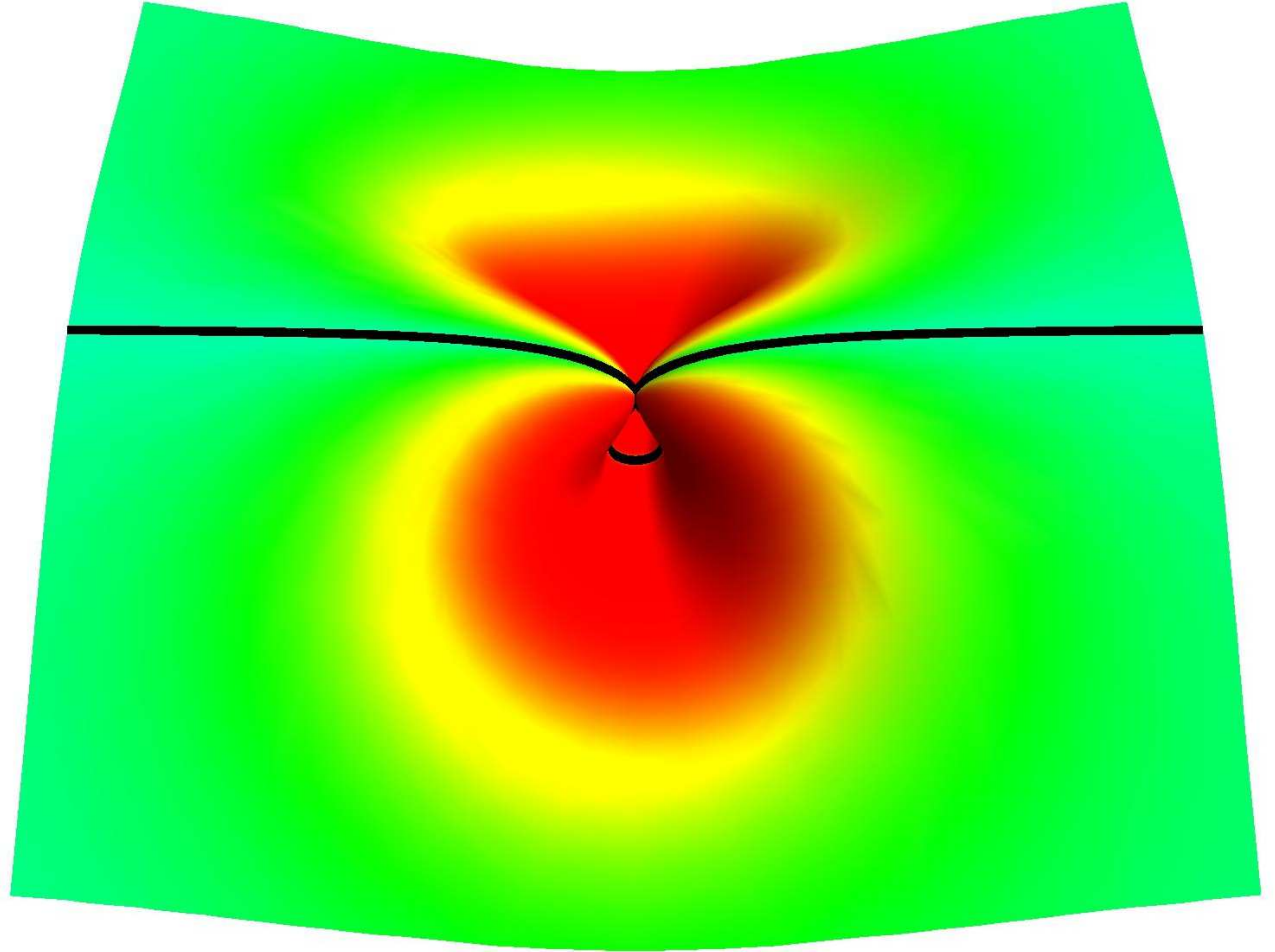}}\\
\includegraphics[scale=0.15]{fig8d.pdf}
\caption{\small (Color online) $\tanh$ of stress distributions in defect. Black
lines represent points where they vanish.}
\label{fig10}
\end{center}
\end{figure}

Using Eq. (\ref{eq:finvgen}), one finds the force per unit length transmitted across a circular
loop with tangent $\widetilde{\bf V}_\parallel$ is given in terms of quantities measured along the
parent catenoid by
\begin{equation}
\tilde{\bf f}_\perp = -\tilde V_{a\, \perp} \tilde{\bf f}^a =
\frac{4 |{\bf X}|^2}{R^4} C_{\perp} {\bf X} \times {\bf V}_\parallel\,.
\end{equation}
A similar identity is obtained for $\widetilde{\bf f}_\parallel$ with $\perp$ replaced by
$\parallel$. The stress is clearly localized near the poles, represented by $|{\bf X}|\to \infty$.
\vskip1pc \noindent
Although the exact expression for the stress tensor is simple to write down, its physical
interpretation is not transparent.  This is facilitated by examining its projections onto the
orthonormal basis adapted to its curvature, $\{\widetilde{\bf V}_\perp, \widetilde{\bf V}_\|,
\widetilde{\bf n}\}$ discussed in Sec. \ref{curvatures}. One has
\begin{equation}
 \widetilde{\bf f}_{\perp} := \widetilde {\sf f}_{\perp \perp}\, \widetilde
{\bf V}_\perp  + \widetilde f_\perp \, \widetilde {\bf n}\,,
\end{equation}
where
$\widetilde {\sf f}_{\perp \perp} \equiv - \widetilde{V}_{\perp\,a}
\widetilde{V}_{\perp\,b} \widetilde{\sf f}^{a b}$ and $\widetilde {f}_{\perp}
\equiv -\widetilde{V}_{\perp\,a} \widetilde{f}^{a}$ are given by
\begin{subequations}
\begin{eqnarray}
\widetilde {\sf f}_{\perp \perp} &=& \frac{4}{R^4}\, C_\perp \, |{\bf X}|^2 \,
{\bf X} \cdot {\bf n}
\,,\\
\widetilde f_\perp &=&  \frac{4 }{R^4}\, C_\perp |{\bf X}|^2\,
{\bf X} \cdot {\bf V}_\perp \,.
\end{eqnarray}
\end{subequations}
The counterpart of $\widetilde {\sf f}_{\perp \perp}$, $\widetilde{\sf f}_{\parallel \parallel}$
satisfies $\widetilde{\sf f}_{\parallel \parallel} = -\widetilde {\sf f}_{\perp \perp}$, a direct
consequence of the traceless character of the tangential stress tensor,  which is itself a
manifestation of the scale invariance of the  energy. The off-diagonal
projection $\widetilde {\sf f}_{\parallel \perp}\equiv \widetilde V_{\parallel\,a} \widetilde
V_{\perp\,b} \widetilde {\sf f}^{a b}$ vanishes in this frame.\footnote{There is, of course, no
shear stress supported by a fluid membrane. A non-diagonal tensor will occur when the tensor is
decomposed with respect to an inappropriate frame.} If the membrane is under tension in the
direction $\widetilde{\bf V}_\perp$, it will be under an equal and opposite compression in the
orthogonal direction $\widetilde{\bf V}_\|$, and vice-versa. The tension and compression will also
assume their maximum and  minimum values along these directions.  $\widetilde{\sf f}_{\perp \perp}$
is represented in Fig. \ref{fig10}(a) for the principal curves $\widetilde{\bf V}_\parallel$
represented in Fig. \ref{fig9}(a). It is proportional to the mean curvature of the defect geometry:
$\widetilde{\sf f}_{\perp \perp}=-(|{\bf X}|/\rho)^2 \widetilde{K}$; thus tangential stresses vanish
wherever  $\tilde K=0$ (cf. Fig. \ref{fig8}(a)). Tension and compression anticorrelate with the sign
of the mean curvature.
\vskip1pc\noindent
Near the poles the defect is under radial tension in response to the external force. Moving away
from the poles,  the dorsal side of the pole remains under radial tension; however,  a transition
from radial tension to compression ($\widetilde{\sf f}_{\perp \perp} < 0$) occurs on descent into
the valley that stretches between the poles.\footnote{This asymmetry in the distribution of stress
under external forces has its counterpart in a familiar daily routine: pulling a T-shirt by the back
of the collar over one's head; whereas tension is established in the shirt along one's back,  it
will be compressed along one's chest.} The radial tension is accompanied by lateral compression,
and  the lateral compression along the valley is accompanied by tension along its
length.\footnote{The distribution of stress along the valley is also consistent with one's
expectations for a surface of approximately constant positive mean curvature.} The membrane is under
rapidly decaying radial tension everywhere far from the defect center. These results are consistent
with those obtained in the quadratic approximation in Sec. \ref{asymptension}.
\vskip1pc\noindent
The projection of the normal stress $\widetilde{f}_\perp$ is represented in Fig. \ref{fig10}(b). It
is negative everywhere, except along $\widetilde{\Gamma}_\parallel$ where it vanishes,
so the normal external force transmitted across these closed curves points in a direction
opposite that of the surface normal (see Fig. \ref{fig6}(c)).
\vskip1pc \noindent
For completeness, the counterpart $\widetilde{f}_\parallel$ normal to the principal curves
$\widetilde{\bf V}_\perp$ is represented in Fig. \ref{fig10}(c).\footnote{\sf The appropriate
Darboux basis, indicated in Fig. \ref{fig9}(b), is given by $\widetilde{\bf t} = - \sign
\widetilde Y\,\widetilde{\bf V}_\perp$ and $\widetilde{\bf l} = \sign \widetilde Y\,\widetilde{\bf
V}_\parallel$.} It is concentrated within the valley region and positive everywhere (directed along
the normal)  to the surface except along $\widetilde{\Gamma}_\perp$, where it vanishes.

\section{Discussion}

In general, the nonlinearity of the underlying geometrical theory makes it difficult to
gather insight into the physical behavior of fluid membranes without resorting to brute force.
The conformal invariance of the bending energy, however, offers access to their nonlinear response--albeit idealized--to external forces, a regime in which one would not have expected to admit an analytical description.
\vskip1pc \noindent
This symmetry has been used to construct a model of a fluid vesicle with a local defect held together by localized external forces. The deformations of the spherical geometry are large at the center
of the defect and they do not lend themselves to a global perturbative description. Indeed, serious shortcomings of the linear Monge representation of the surface have been exposed where results, at best, are misleading:
Small gradients may have large curvatures lurking at their center.
\vskip1pc \noindent
More abstractly, the conformal symmetry implies a duality between the weak far-field description in
one geometry--the catenoid--and the strong near field behavior in another--the defect. Whereas the
catenoid is a stress--free state, its dual exhibits a localized distribution of stress, supported by
external sources, with curvature singularities at its center.  At a technical level, this duality
establishes a link between harmonic functions which describe the local behavior of minimal surfaces
and biharmonic counterparts which describe an unconstrained fluid membrane. More complex defects could, in principle,
be constructed using other minimal surfaces.

\vskip1pc\noindent
For the purpose of containing the narrative,
we have focused almost completely on symmetric defect states. Theses are not the only states of
interest. Indeed free passage into states which do not exhibit this symmetry is possible due to the
existence of the zero mode.  One way to curtail this freedom is to introduce additional
constraints.  One possibility, in the spirit of the bilayer couple model,  is to impose a
constraint on the area difference. This would involve adding to the Hamiltonian a term, $\beta (\int
dA\, K - M_0)$ \cite{SvetinaZeks}. Deformations will now be constrained to also satisfy $\int dA \,
K_G \Phi=0$, which will lift this degeneracy.  Interestingly, there will generally be no mirror
symmetric state consistent with a given value of $M_0$.  Indeed, $M_0$ may also be negative, a
possibility that was never even contemplated in the original axisymmetric analysis of the bilayer
couple model where the parameter space was truncated at $M_0=0$. It would also be interesting to
understand how defects on vesicles behave when conformal symmetry is broken: Accommodating an area
reservoir, a pressure difference, or a spontaneous curvature  would do this; an external force dipole
holding the two points a finite, nonvanishing, Euclidean distance apart would also.  While it is
not going to be pretty, it should be straightforward to develop perturbation theory around one of
the equilibrium states that we have described.

\section*{Acknowledgments}

We have benefited from conversations with Pavel Castro-Villarreal, Markus Deserno, Martin Mueller,
and Francisco Solis. Support from DGAPA PAPIIT Grant No. IN114510-3 and CONACyT Grant No. 180901 is
acknowledged. P.V.M. is also grateful to UAM-Cuajimalpa for financial support.

\begin{appendix}

\setcounter{equation}{0}
\renewcommand{\thesection}{Appendix \Alph{section}}
\renewcommand{\thesubsection}{A. \arabic{subsection}}
\renewcommand{\theequation}{A.\arabic{equation}}

\section{Conformal Invariance and Constraints} \label{appCIC}

Suppose that the surface is described by its embedding into Euclidean space, $(u^1,
u^2)\to {\bf X}(u^1,u^2)$. Consider the problem of minimizing the energy of a closed vesicle with
two of its points brought into contact (say, ${\bf X}_1$ and ${\bf X}_2$) subject to the constraints that both the area and the volume are fixed. The contact constraint is incorporated formally into the variational principle by introducing a new Lagrange multiplier, ${\bf F}$, to tie the points,
\begin{equation}
H[{\bf X}] = H_{B} [{\bf X}] + \sigma \, (A [{\bf X}]- A_0) + P\, (V [{\bf X}]-V_0) + {\bf F}\cdot ( {\bf X}_1 - {\bf X}_2)\,,
\end{equation}
where $H_B$ is the bending energy of the membrane given by the sum of the first two
terms in Eq. (\ref{CHenergy}), $H_B[{\bf X}] = \kappa H_1[{\bf X}] + \bar\kappa H_{GB} [{\bf X}]$.
Consider the change in $H$ under the conformal transformation \cite{Seifert},
\begin{equation}
\label{sct}
\delta {\bf X}= \lambda \mathbf{ X} + \mathbf{ X}^ 2 \, \mathbf{c} -
 2 (\mathbf{X}\cdot \mathbf{c}) \,\mathbf{X}\,,
\end{equation}
where $\lambda$ is a constant, and $\mathbf{c}$ is a constant vector.  One has
\begin{equation}
\delta H[{\bf X}] = \sigma \, \delta A + P\, \delta V \,.
\end{equation}
We have used the conformal invariance of the bending energy as well as the contact constraint.
In an equilibrium state, $\delta H$ must vanish, whether  or not $\delta \mathbf{X}$ preserves the area or the volume. In particular,  one can tune the four parameters $\lambda$ and $\mathbf{c}$ in either of two independent ways: i) the area is preserved ($\delta A =0$) but the volume is not ($\delta V\ne 0$) or ii) the volume is preserved ($\delta V=0$) but the area is not ($\delta A\ne 0$). Such variations exist; together,  they imply that the two multipliers
vanish in the equilibrium end state.
\footnote{Under a rescaling of the surface, $\delta {\bf X}= \lambda {\bf X}$, one finds the weaker condition, $2\sigma A + 3 P V=0$ in equilibrium states.}
\footnote{In the absence of the contact constraint, the only equilibria are
axially symmetric with mirror symmetry with respect to a plane orthogonal to the axis of symmetry.
This is a consequence of the fact  that, when the vesicle is  closed,  $\delta A = 2\lambda A -
4 \mathbf{c} \cdot \int dA\,\mathbf{X}$, and $\delta V = 3 \lambda V - 2\mathbf{c} \cdot \int dA\,(\mathbf{n}\cdot\mathbf{X})(\mathbf{X}\cdot\mathbf{c})$, where $\mathbf{n}$ is the surface normal \cite{Seifert}.
One of the two integrals involving $\mathbf{c}$ can always be set to zero by locating the
appropriate center of mass at the origin.  In general, however, the second will not vanish.} 
Had the two points been held a fixed nonvanishing distance apart, as in any of the
intermediate states in the sequence, this distance would set a new scale that competes with the one
established by the size of the membrane. One then faces a more difficult problem without conformal
invariance to direct us towards a solution and about which we have nothing to say. It should be
emphasized that vanishing $\sigma$ does not imply that the tension in the vesicle vanishes. The
forces bringing the points together establish stress in the vesicle.

\setcounter{equation}{0}
\renewcommand{\thesection}{Appendix \Alph{section}}
\renewcommand{\thesubsection}{B. \arabic{subsection}}
\renewcommand{\theequation}{B.\arabic{equation}}

\section{Transformation of the geometry under inversion} \label{appinvgeom}

Let ${\bf e}_a$ be the two tangent vectors to the surface adapted to the
parametrization by $Z$ and $\varphi$, The corresponding tangent vectors on
the defect geometry are given by
\begin{equation} \label{eq:einvert}
\widetilde{\bf e}_a  = \left(\frac{R}{|{\bf X}|}\right)^2\,{\sf R} \,{\bf
e}_a\,,
\end{equation}
where the linear mapping
${\sf R}= \mathbbm{1} - 2 \hat{\bf X} \otimes \hat{\bf X}$, represents a
reflection of surface points in the plane passing through the origin and
orthogonal to ${\bf X}$. $\hat{\bf X}$ denotes the corresponding unit vector.
\vskip1pc \noindent
The metric tensor induced on the surface are given by $g_{ab} = {\bf e}_a\cdot {\bf e}_b$. The
metric under inversion is related to $g_{ab}$ by
\begin{equation} \label{eq:gabinv}
g_{ab}\to \widetilde{g}_{ab} = \frac{R^4}{|{\bf X}|^4}\, \,g_{ab}\,.
\end{equation}
Let ${\bf n}$ be the unit normal to the catenoid and $K_{ab}  = {\bf e}_a\cdot
\partial_b {\bf n}$ the extrinsic curvature tensor on this surface. The
counterparts on the defect are given by $\widetilde {\bf n}= - {\sf R}\, {\bf
n}$,\footnote{The minus sign preserves the orientation of the adapted
basis.} and
\begin{equation}  \label{eq:Ktr}
\widetilde{K}_{ab} = -\frac{R^2}{|{\bf X}|^2}\, \left(K_{ab} -\frac{2}{|{\bf X}|^2}\,
{\bf X}  \cdot{\bf n} \, g_{ab} \right)\,.
\end{equation}
It involves only the two tensors, $g_{ab}$ and $K_{ab}$. A derivation is
provided in Ref. \cite{Trinoid}.  Let $C_\parallel$ and $C_\perp$ denote
the principal curvatures and ${\bf V}_\parallel= V_\parallel^a {\bf e}_a$,
${\bf V}_\perp = V_\perp^a {\bf e}_a$ the corresponding directions on the
catenoid, so $K^a_{\phantom{a}b} V_i^b = C_i V_i^a$, $i=\parallel, \perp$. Then
$\widetilde{K}^a_{\phantom{a}b} V_i^b = \widetilde{C}_i V_i^a$,
where
\begin{equation} \label{eq:cbarc}
\widetilde{C}_i = - \frac{|{\bf X}|^2}{R^2}\,\left( C_i - \frac{2}{|{\bf
X}|^2} {\bf X}\cdot {\bf n} \right)\,.
\end{equation}
The corresponding principal directions are the normalized images of ${\bf V}_\parallel$ and ${\bf V}_\perp$ on the transformed surface:
\begin{equation}
\widetilde {\bf V}_i = \widetilde V^a_i \widetilde {\bf e}_a =
{\sf R} \, {\bf V}_i\,.
\end{equation}
These directions are thus appropriate rotations of their counterparts on the original surface and their components
transform as $\widetilde V^a_i = |{\bf X}|^2/R^2 V^a_i$.
\vskip1pc \noindent
As a consequence of the relation (\ref{eq:cbarc}), the difference of the two
principal curvatures under inversion is preserved, modulo a scaling by the
distance function and a change of sign,
\begin{equation} \label{eq:difinvcurv}
\widetilde C_\parallel - \widetilde C_\perp = - \frac{|{\bf X}|^2}{R^2}\, (C_\parallel - C_\perp)\,.
\end{equation}
In particular, on an  inverted minimal surface (with a
vanishing mean curvature, $K=0$), one has $\widetilde{K}  =  4/ {R^2} \, {\bf
X} \cdot{\bf n}$.

\setcounter{equation}{0}
\renewcommand{\thesection}{Appendix \Alph{section}}
\renewcommand{\thesubsection}{C. \arabic{subsection}}
\renewcommand{\theequation}{C.\arabic{equation}}

\section{Gauss-Bonnet Energy} \label{appGB}

Here the transformation of the Gauss-Bonnet energy under inversion is examined. Specifically it
will be shown that the change under inversion is a sum of boundary terms. To reduce the burden of
notation, consider the inversion  in the unit sphere. One has, using
Eq. (\ref{eq:cbarc}), with $\tilde {\cal K}_G =\tilde C_1 \tilde C_2$,
\begin{equation}
\int d\tilde A\, \tilde {\cal K}_G = \int dA \,{\cal K}_G - 2\int dA\,
\left(\frac{{\bf X}\cdot {\bf n}}{|{\bf X}|^{2}}\right)\, K + 4\,
\int dA\, \left(\frac{{\bf X}\cdot {\bf n}}{|{\bf X}|^2}\right)^2\,.
\label{eq:id2}
\end{equation}
However,
\begin{eqnarray}
\int dA\, \left( \frac{{\bf X}\cdot {\bf n}}{|{\bf X}|^{2}}\right)\, K
&=& - \int dA \, \left(\frac{{\bf X}\cdot \nabla^2 {\bf X}}{|{\bf X}| ^{2}}\right)
\nonumber\\
 &=& 2 \int dA \, \left(\frac{{\bf X}\cdot {\bf n}}{|{\bf X}|^{2}}\right)^2
- \frac{1}{2} \int dA\, \nabla^2 \ln \,|{\bf X}|^2 \,. \label{eq:id3}
\end{eqnarray}
Thus,
\begin{equation}
\int d\widetilde A\, \widetilde {\cal K}_G = \int dA \,{\cal K}_G -  \int d \widetilde
A\, \widetilde {\nabla^2}  \ln \,|\widetilde{\bf X}|^2 \,. \label{eq:id4}
\end{equation}
In any axially symmetric geometry, the Gauss-Bonnet energy itself can be cast explicitly as a
boundary term. To see this note that $C_1= \dot \Theta$, and $C_2= \sin\Theta /R$ where $R$ is the polar radius,
$\Theta$ is the turning angle that the tangent vector to the generating curve makes
with the radial direction, and the dot represents differentiation with respect to $\ell$, the
arc-length along the meridian \cite{Gray}.
Replacing $\ell$ by $\Theta$ as variable of integration, one obtains $H_{GB}$  as a difference of
cosines,
\begin{eqnarray}
H_{GB}  = 2\pi \int^{\infty}_{0} d \ell \,
R \, \dot{\Theta} \left(\ell\right)~\frac{\sin\Theta\left(\ell\right)}{R} = 2\pi \,(\cos
\Theta_1-\cos \Theta_2)\,.
\end{eqnarray}
The Gaussian energy of a surface of revolution depends only on the boundary values $\Theta_1$ and
$\Theta_2$ of $\Theta$. For a catenoid ($\Theta_{1,2} = \pi,0$), $H_{GB}= - 4\pi$;
whereas $H_{GB} = 4\pi$ for its axially symmetric inverted counterpart of spherical topology
($\Theta_{1,2} = 0, \pi$) with a well-defined tangent plane at its
poles.\footnote{The sphere of inversion is not centered on the surface, otherwise the resulting
surface has a planar topology, and in consequence, $H_{GB}=0$.}
\vskip1pc\noindent
This discrepancy is consistent with  Eq. (\ref{eq:id4}). To see this, note that the
boundary term on the right in this equation is a sum of two terms of the form
\begin{equation}
 2 \int d\widetilde s\, \frac{\widetilde {\bf l}\cdot \widetilde {\bf X}}{ |\widetilde {\bf
X}|^2}= - 4\pi\,,
\end{equation}
one for each pole.
\end{appendix}


\begin{thebibliography}{30}

\bibitem{canhamHelfrich} P. Canham, {\it J. Theor. Biol.} {\bf 26}, 61 (1970); W. Helfrich,
{\it Z. Naturforsch. C} {\bf 28}, 693 (1973).

\bibitem{Willmore} T. J. Willmore, {\it Total Curvature in Riemannian Geometry} (Ellis Horwood, Chichester, 1982).

\bibitem{Polyakov} A. M. Polyakov, {\it Gauge Fields and Strings} (Harwood
Academic Publishers, Newark, NJ, 1987).

\bibitem{Seifert} U. Seifert, {\it J. Phys. A:Math and Gen.}{\bf 24}, L573 (1991);
F. J\"ulicher, U. Seifert and R. Lipowsky {\it Phys. Rev. Lett.} {\bf 71}, 452 (1993).

\bibitem{guvencastro} P. Castro-Villarreal and J. Guven, {\it J. Phys. A: Math. and Theor.} {\bf 40}, 4273 (2007).

\bibitem{invcat} P. Castro-Villarreal and J. Guven, {\it Phys. Rev. E} {\bf 76}, 011922 (2007).

\bibitem{Trinoid} P. Castro-Villarreal, J. Guven and P. V\'azquez-Montejo, in preparation.

\bibitem{SvetinaZeks} S. Svetina, and B. Žekš, {\it Eur. Biophys. J.}  {\bf 17}, 101 (1989).

\bibitem {SeifertBerndlLipowsky} U. Seifert, K. Berndl, and R. Lipowsky, {\it Phys. Rev. A} {\bf  44}, 1182
(1991).

\bibitem{Gray} A. Gray, {\it Modern Differential Geometry of Curves and
Surfaces with Mathematica} (Chapman \& Hall/CRC, London, 2006).

\bibitem{Fomenko} A. T. Fomenko and A. A. Tuzhilin, {\it Elements of the
Geometry and Topology of Minimal Surfaces in  Three-Dimensional Space} (Translations of Mathematical
Monographs Vol. 93, American Mathematical Society, Providence, RI, 1991).

\bibitem{Defo} R. Capovilla, J. Guven, and J. A. Santiago,
{\it J. Phys A: Math. and Gen.}  {\bf 36}, 6281 (2003).

\bibitem{Stress} R. Capovilla and J. Guven, {\it J. Phys. A: Math. and Gen.} {\bf 35}, 6233 (2002).

\bibitem{Auxiliary} J. Guven, {\it J. Phys. A: Math. and Gen.} {\bf 38}, L313 (2004).

\bibitem{Jenkins} J. T. Jenkins, {\it SIAM J. Appl. Math.} {\bf 32}, 755 (1977).

\bibitem{Steigman} D. J. Steigmann, {\it Arch. Ration. Mech. Anal.} {\bf 150}, 127 (1999).

\bibitem{Lomholt} M. A. Lomholt and L. Miao, {\it J. Phys. A.} {\bf 39}, 10323 (2006).

\bibitem{MDG} M. M. M\"uller, M. Deserno and J. Guven, {\it Euro. Phys. Lett.} {\bf 69} (2005), 482;
{\it Phys. Rev. E} {\bf 72}, 061407 (2005).

\bibitem{Fournier} J. B. Fournier, {\it Soft Matter} {\bf 3}, 883 (2007).

\bibitem{ConfJ} J. Guven, {\it J. Phys. A: Math and Gen.} {\bf 38}, 7943 (2005).

\bibitem{Laplace} J. Guven, {\it J. Phys. A: Math and Gen.} {\bf 39}, 3771 (2006).

\end{thebibliography}
\end{document}